\documentclass[aps,prd,10pt,twocolumn,superscriptaddress,showpacs,footinbib]{revtex4-1}

\usepackage{graphicx}
\usepackage{amsfonts,amsmath,amssymb,bm,bbm}
\usepackage{color}
\usepackage{setspace}
\usepackage{booktabs, tabularx}
\usepackage{multirow}

\usepackage{hyperref}
\hypersetup{
    pdfnewwindow=true,      % links in new window
    colorlinks=true,       % false: boxed links; true: colored links
    linkcolor=blue,          % color of internal links
    citecolor=blue,        % color of links to bibliography
    filecolor=blue,      % color of file links
    urlcolor=blue        % color of external links
}

\begin{document}

\title{Lensing of fast radio bursts: future constraints on primordial black hole density with an extended mass function and a new probe of exotic compact fermion and boson stars}

\author{Ranjan Laha}
\affiliation{PRISMA Cluster of Excellence and
             Mainz Institute for Theoretical Physics,
             Johannes Gutenberg-Universit\"{a}t Mainz, 55099 Mainz, Germany \\
{\tt  \href{mailto:ranjan.laha@cern.ch}{ranjan.laha@cern.ch}}
{\tt \footnotesize \href{http://orcid.org/0000-0001-7104-5730}{0000-0001-7104-5730} \smallskip}}

\date{\today}

\begin{abstract}
The discovery of gravitational waves from binary black hole mergers has renewed interest in primordial black holes forming a part of the dark matter density of our Universe.  Various tests have been proposed to test this hypothesis.  One of the cleanest tests is the lensing of fast radio bursts.  In this situation, the presence of a compact object near the line of sight produces two images of the radio burst.  If the images are sufficiently separated in time, this technique can constrain the presence of primordial black holes.  One can also try to detect the lensed image of the mini-bursts within the main burst.  We show that this technique can produce the leading constraints over a wide range in lens masses $\gtrsim$ 2 $M_\odot$ if the primordial black holes follow a single mass distribution.  Even if the primordial black holes have an extended mass distribution, the constraints that can be derived from lensing of fast radio bursts will be the most constraining over wide ranges of the parameter space.  We show that this technique can probe exotic compact boson stars and fermion stars made up of beyond the Standard Model particles.  This search strategy is competitive and can provide the leading constraints on parts of the particle physics parameter space when compared with gravitational wave observations.
\end{abstract}

\maketitle

%%%%%%%%%%%%%%%%%%%%%%%%%%%%%%%%%%%%%%%%%%%%%%%%%%%
%%%%%%%%%%%%%%%%%%%%%%%%%%%%%%%%%%%%%%%%%%%%%%%%%%%

\section{Introduction}
\label{sec:introduction}

The direct detection of gravitational waves has heralded a new era in understanding of our Universe.  The discovery of black hole mergers and neutron star mergers and its associated gravitational waveforms have confirmed numerous predictions and unveiled answers to some of the major questions in astrophysics\,\cite{Abbott:2016blz, TheLIGOScientific:2017qsa, LIGOScientific:2018jsj}.  These events have also been used to test various modifications of gravity and Standard Model extensions.  

Almost immediately after the discovery of the first binary black hole merger, it was suggested that these astrophysical objects might be primordial in nature and contribute to the dark matter density of our Universe\,\cite{Bird:2016dcv, Cholis:2016kqi, Raccanelli:2016cud, Munoz:2016tmg, Clesse:2016vqa, Sasaki:2016jop, Kashlinsky:2016sdv}.  Primordial black holes (PBHs) are produced in the Universe due to enhanced density perturbations and their existence have been speculated for a long time\,\cite{1966AZh....43..758Z, Hawking:1971ei, Carr:1974nx, Meszaros:1975ef, Carr:1975qj, Khlopov:2008qy}.  Numerous constraints exist on the present density of PBHs in our Universe.  These constraints include searching for  capture of PBHs by compact objects\,\cite{Takhistov:2017nmt}, lensing searches\,\cite{Niikura:2017zjd, Griest:2013esa, Allsman:2000kg, Tisserand:2006zx, Wyrzykowski:2011tr, Zumalacarregui:2017qqd, Mediavilla:2017bok}, dynamical effects  on ultra-faint dwarf galaxies\,\cite{Brandt:2016aco, Koushiappas:2017chw}, orbital dynamics\,\cite{Monroy-Rodriguez:2014ula}, non-observations of stochastic gravitational waves\,\cite{Wang:2016ana, Clesse:2016ajp}, and effects of accretion onto PBHs via cosmic microwave background observations\,\cite{Ali-Haimoud:2016mbv, Poulin:2017bwe, Bernal:2017vvn}.  Besides these searches which are applicable for macroscopic compact objects, traditional techniques in dark matter indirect detection like searching for gamma-rays, charged cosmic rays\,\cite{Laha:2012fg, Ng:2013xha, Chowdhury:2016bxs}, or neutrinos\,\cite{Dasgupta:2012bd, Murase:2015gea} can also be used to search for PBHs if they are sufficiently light and produce these particles via evaporation\,\cite{Boudaud:2018hqb, Laha:2019ssq, Dasgupta:2019cae, Clark:2016nst, Laha:2020ivk}.  Some of the constraints studied in the literature are controversial\,\cite{Gaggero:2016dpq, Hektor:2018rul, Manshanden:2018tze}, some require a detailed understanding of the merger rate\,\cite{Ballesteros:2018swv}, while others require a detailed understanding of wave optics and source size effects\,\cite{Katz:2018zrn}, and the effect of clustering\,\cite{Garcia-Bellido:2017xvr, Ali-Haimoud:2018dau, Desjacques:2018wuu, Belotsky:2018wph, Bringmann:2018mxj}. There have also been studies regarding future constraints on primordial black holes involving fast radio bursts (FRBs), pulsar timing, 21 cm signals, galaxy clustering, gravitational waves, gamma-ray bursts, and X-ray pulsars\,\cite{Zheng:2014rpa, Munoz:2016tmg, Wang:2018ydd, Eichler:2017eid, Schutz:2016khr, Hektor:2018qqw, Clark:2018ghm, Raccanelli:2017xee, Bartolo:2018rku, Ji:2018rvg, Katz:2018zrn, Bai:2018bej}.  

FRBs are intense radio pulses on the sky which have $\sim \mathcal{O}$(1 -- 10 ms) duration.  These puzzling astrophysical transients were first discovered in 2007\,\cite{Lorimer:2007qn}, and till date $\sim$ $\mathcal{O}$(100)  FRBs have been discovered\,\cite{Petroff:2016tcr, 2017arXiv171008155P}.  Temporal structures within the main burst having $\sim \mathcal{O}$(10 $\mu$s) duration have also been observed\,\cite{Farah:2018buz, Hessels:2018mvq}.  The dispersion measure of these radio waves, which quantify the amount of electrons in the line of sight from the source to the Earth, signifies that these objects are extragalactic in nature\,\cite{Niino:2018xze}, and there is direct evidence for the extragalactic nature of FRB121102\,\cite{Chatterjee:2017dqg, Tendulkar:2017vuq}.  The radio bursts from FRB121102 and FRB 180814.J0422+73 repeat, however, the pulses are not periodic\,\cite{Spitler:2016dmz, Scholz:2017kwy, Amiri:2019bjk}.  The detection rate of FRBs depends on the observing frequency and at frequencies between 400 MHz and 800 MHz, it can be as large as 1 per hour\,\cite{Connor:2016rhf, Lawrence:2016qzx, Chawla:2017wdz}.  Recently, during a pre-commissioning phase of the telescope, CHIME discovered 13 FRBs demonstrating its potential to be one of the foremost FRB discovering telescopes\,\cite{Amiri:2019qbv}.  A number of review papers summarizing the properties of FRBs exist in the literature\,\cite{Platts:2018hiy, Pen:2018ilo, Lorimer:2018rwi, Keane:2018jqo, Kulkarni:2018ola, 2018NatAs...2..192M, Macquart:2018fhn, Caleb:2018ygr, Burke-Spolaor:2018xoa, Popov:2018hkz, Katz:2018xiu, Popov:2018wei, Petroff:2017pny}.

Due to their large event rate, small temporal duration, and cosmological origin, it has already been realized that FRBs can be used as cosmological probes to test various fundamental principles\,\cite{Zhou:2014yta, Gao:2014iva, Wu:2016brq, Wei:2015hwd, Li:2017mek, Munoz:2018mll}.  In this paper, we concentrate on lensing of fast radio bursts\,\cite{Zheng:2014rpa, Munoz:2016tmg, Wang:2018ydd}.  In this situation, the presence of a sufficiently compact object near the line of the sight between the source and the observer causes multiple images of the source.  The resulting images are separated in space and time.  Due to the narrow temporal duration of the main burst, it is possible that the arrival time of the various images do not overlap with each other.  This phenomenon of time delay between various lensed images has been confirmed in the case of a supernova (acting as a source) and a galaxy (acting as a lens)\,\cite{Kelly:2015xvu}.  Given that the most FRBs (notable exceptions include FRB121102 and FRB 180814.J0422+73) do not repeat, this can be a very clean test of the existence of massive compact objects.  The sample of repeating FRBs can also be used to constrain compact dark objects:\,the lensed image appears after a predicted time interval and with a predicted magnification factor.  On the other hand, the bursts which repeat (due to non-lensing astrophysical reasons) do not follow this temporal and magnification rules.  It is expected that upcoming radio telescopes will detect a large number of FRBs, and it is possible to search for the lensing signal.  In particular, CHIME is expected to detect 2 -- 42 FRBs per day, and during its operational duration (atleast 3 years), it can detect a total of $\sim$ 2000 --- 46000 FRBs\,\cite{2018ApJ...863...48C}.  It will be shown that such a large number of FRBs will permit an extremely restrictive constraint on the abundance of PBHs.  In addition, there are several other telescopes (like HIRAX\,\cite{Weltman:2017sxz}, APERTIF\,\cite{Maan:2017uts}, UTMOST\,\cite{2017PASA...34...45B}, Ooty Wide Field Array\,\cite{Bhattacharyya:2018alq}, and the upgraded Giant Metrewave Radio Telescope\,\cite{Bhattacharyya:2018alq}) which will revolutionize the field of FRBs.

Building on the study in Ref.\,\cite{Munoz:2016tmg}, we will calculate the future constraints on PBH density assuming an extended power law mass distribution.  We explicitly show the results of a future PBH search if the FRB distribution follows the star-formation rate.  The parameter space where FRB lensing can probe PBH density can also be searched for by other techniques, and these have already ruled out PBHs as the dominant dark matter component in some of these regions for different kinds of mass distribution\,\cite{Carr:2020gox, Carr:2020xqk, Green:2016xgy, Kuhnel:2017pwq, Bellomo:2017zsr, Lehmann:2018ejc}.  However, it is important to remember that detection of PBHs (even forming an extremely small dark matter density) has tremendous implications for several models of the early Universe.  As such, FRB lensing can also be seen as an important probe for numerous inflationary models\,\cite{Garcia-Bellido:2017mdw, Inomata:2017vxo}.  In addition, we will show that FRB lensing can also probe exotic boson stars and fermion stars.  These compact objects can arise in many well-motivated particle physics models, and we will demonstrate the part of the particle physics parameter space which can be probed by FRB lensing.

We review the theory of FRB lensing and compact objects in Sec.\,\ref{sec:lensing and exotic compact objects}.  We will detail the necessary formula and study how compact an object needs to be for our lensing formalism to be valid.  We will mention our results in detail in Sec.\,\ref{sec:results}, where we will present future constraints on PBH density from FRB lensing for single mass distribution and extended mass distributions for two different redshift distributions of FRBs, and outline the particle physics parameter space which FRB lensing will probe provided exotic compact boson stars and fermion stars are abundant in the Universe.  We conclude in Sec.\,\ref{sec:conclusions}.

%%%%%%%%%%%%%%%%%%%%%%%%%%%%%%%%%%%%%%%%%%%%%%%%%%%
%%%%%%%%%%%%%%%%%%%%%%%%%%%%%%%%%%%%%%%%%%%%%%%%%%%

\section{Lensing of Fast Radio Bursts and Exotic Compact Objects}
\label{sec:lensing and exotic compact objects}

The theory of lensing is a well-studied phenomenon in general relativity and there are extensive observations which validate the underlying physics.  Lensing is used as an observational tool in astrophysics to probe objects which are too faint to detect by conventional means.  It has been realized long ago that lensing can probe exotic compact objects in our Universe if these objects are sufficiently numerous.  We first review the necessary formalism for lensing of FRBs and then various particle physics models which can give rise to exotic compact objects in the form of boson stars or fermion stars.

%%%%%%%%%%%%%%%%%%%%%%%%%%%%%%%%%%%%%%%%%%%%%%%%%%%

\subsection{Lensing of fast radio bursts}
\label{sec:FRB lensing}

We consider a scenario where a fast radio burst acts as the source and is at an angular diameter distance $D_S$ from the observer.  We assume that a single compact object, of mass $M_L,$ is in between the source and the observer and acts as the lens.  The angular diameter distance between the lens and the observer and between the source and the lens is denoted by $D_L$ and $D_{LS}$, respectively.  If the source and the lens are situated at redshifts $z_S$ and $z_L$, respectively, then the angular diameter distance between them is\,\cite{Hogg:1999ad}
\begin{eqnarray}
D_{LS} = \dfrac{1}{1+z_S} \, \left[\chi(z_S) - \chi(z_L)\right] \, ,
\label{eq:angular diameter distance between source and lens}
\end{eqnarray}  
where $\chi(z_S)$ and $\chi(z_L)$ denote the co-moving distance to the source and the lens, respectively.  We assume that the energy density due to the curvature of space is 0 throughout the paper.  For the other relevant cosmological parameters, we take the best fit values from PDG\,\cite{Tanabashi:2018oca}.

For a point lens, the Einstein radius is given by\,\cite{Narayan:1996ba, Takahashi:2003ix, Paolis:2016roy}
\begin{eqnarray}
\theta_E = 2 \sqrt{\dfrac{G_N M_L D_{LS}}{c^2  D_S D_L}}\,,
\label{eq:Einstein radius}
\end{eqnarray}  
where $G_N$ and $c$ denote the Newton's gravitational constant and the speed of light respectively.  Due to our assumption that we only consider point lens, the radius of the compact lens, $R_L$, is constrained to be much smaller than the the Einstein radius multiplied by the angular diameter distance of the lens.  This can be expressed as
\begin{eqnarray}
R_L \ll D_L (z_L) \times 2 \sqrt{\dfrac{G_N M_L}{c^2} \, \dfrac{D_{\rm LS}(z_S, z_L)}{D_L(z_L) \, D_S(z_S)}} \, .
\label{eq:size of lens for point mass approximation}
\end{eqnarray}
In Sec.\,\ref{sec:results}, we will demonstrate this constraint on the lens size assuming various different inputs.

The point lens produces two images of the source and these are temporally separated by
\begin{eqnarray}
&&\Delta t = 4 \dfrac{G_N M_L}{c^3} \, (1 + z_L) \nonumber \\
&\times& \Bigg[\dfrac{y}{2} \sqrt{y^2 + 4} + {\rm log} \Bigg(1 + \dfrac{2y}{\sqrt{4 + y^2} - y} \Bigg) \Bigg] \, ,
\label{eq:time delay between two images}
\end{eqnarray}
where $y$ is the normalized impact parameter and is defined as the ratio of the angular impact parameter to the angular Einstein radius.  The lensing cross section due to a point lens is given by
\begin{eqnarray}
&&\sigma (M_L, z_L) = \dfrac{4\pi G_N M_L D_L D_{LS}}{c^2 D_S} \nonumber\\
&\times& \Big[y_{\rm max}^2(\overline{R}_f) - y_{\rm min}^2(M_L, \,z_L) \Big] \,,
\label{eq:lensing cross section}
\end{eqnarray}
where the maximum and minimum values of the impact parameters are denoted by $y_{\rm max}$ and $y_{\rm min}$, respectively.  The maximum value of impact parameter can be found by requiring that the magnification of the two lensed images is greater than some reference value $\overline{R}_f$, 
\begin{eqnarray}
1 + \dfrac{2 y \sqrt{y^2 + 4}}{y^2 + 2 - y \sqrt{y^2 + 4}} \leq \, \overline{R}_f \,.
\label{eq:ymax equation}
\end{eqnarray}
This permits an analytical expression of $y_{\rm max}$,
\begin{eqnarray}
y_{\rm max} (\overline{R}_f) = \sqrt{\dfrac{1+\overline{R}_f}{\overline{R}_f^{0.5}} - 2} \,.
\label{eq:ymax expression}
\end{eqnarray}
Following Ref.\,\cite{Munoz:2016tmg}, we take $\overline{R}_f$ = 5 for cases when we study lensing of the whole burst.  When we consider lensing of the sub-bursts, we take $\overline{R}_f$ = 2.5 and 5.  

Assuming that the time delay is greater than a critical time, $\overline{\Delta t}$, one can numerically estimate the value of  $y_{\rm min}(M_L, \,z_L)$ from Eq.\,\ref{eq:time delay between two images}.  This critical time needs to be large enough so that it takes into account the intrinsic broadening of the spectrum due to dispersion of radio waves during propagation.  Depending on the width of the burst, this can have different values for different bursts.  Given that smaller time structures have been observed inside the main burst of some FRBs, we will also show our results for critical values which are smaller than the total temporal duration of the main burst.  In the latter case, one can aim to lens the mini-bursts within the main burst and this will permit constraints on lower lens masses.  In a realistic case, the duration of the burst will be different for different bursts.  As per frbcat.org (date accessed December 13, 2018), the burst duration of FRBs varies from 0.35 ms to 21 ms.  FRB170827 has also shown narrower temporal components down to $\sim$ 30 $\mu$s.  If such narrow time structures exist in other bursts, then there is the possibility that these narrow bursts can also be lensed.  In order to cover this wide range and for simplicity, we consider four different critical times, $\overline{\Delta t}$ = 0.1 ms, 0.3 ms, 1 ms, and 3 ms. 

For a single source, the optical depth for lensing due to a single PBH or exotic compact object is
\begin{eqnarray}
\tau (M_L, z_s) &=& \int_0 ^{z_s} \, d\chi(z_L) \, (1 + z_L)^2 \, n_L \, \sigma (M_L, z_L) \nonumber\\
&=& \dfrac{3}{2} f_{\rm DM} \, \Omega_c \, \int_0 ^{z_s} dz_L \, \dfrac{H_0^2}{c \, H(z_L)} \dfrac{D_L D_{LS}}{D_S} \, (1+z_L)^2 \nonumber\\
&\times& \Big[y_{\rm max}^2(\overline{R}_f) - y_{\rm min}^2(M_L, \,z_L) \Big]\,,
\label{eq:optical depth eqn 1}
\end{eqnarray}
where $n_L$ is the co-moving number density of the lens, $H(z_L)$ is the Hubble function at $z_L$, $f_{DM}$ represents the fraction of dark matter which is in the form of the compact objects acting as the lens, and $\Omega_c$ is the present density of dark matter. 

In order to find the total lensing optical depth, one needs to convolve the optical depth in Eq.\,\ref{eq:optical depth eqn 1} with the redshift distribution of FRBs.  Given the low number of FRBs, the redshift distribution is very uncertain and is an active field of research\,\cite{Deng:2018tjv}.  We will follow the two redshift distributions as studied in Ref.\,\cite{Munoz:2016tmg}: constant-density redshift distribution and star-formation redshift distribution.  As we will see, the constant-density redshift distribution and the star-formation redshift distribution give a conservative and an optimistic result, respectively.

The constant-density redshift distribution can be described as\,\cite{Oppermann:2016mzk, Munoz:2016tmg}
\begin{eqnarray}
N_{\rm const}(z) = \mathcal{N}_{\rm const} \, \dfrac{\chi^2 (z) \, e^{- d_L^2 (z)/[2 d_L^2 (z_{\rm cut})] }}{(1+z)H(z)} \,,
\label{eq:constant-density redshift distribution}
\end{eqnarray}
where $\mathcal{N}_{\rm const}$ is a constant such that $\int dz \, c \, N_{\rm const} (z) = 1$, $d_L (z)$ is the luminosity distance to $z$, and $z_{\rm cut}$ represents the cutoff in the FRB redshift distribution.  The star-formation redshift distribution can be described as\,\cite{Caleb:2015uuk, Munoz:2016tmg}
\begin{eqnarray}
N_{\rm SFH}(z) = \mathcal{N}_{\rm SFH} \, \dfrac{\dot{\rho}_*(z) \chi^2 (z) \, e^{- d_L^2 (z)/[2 d_L^2 (z_{\rm cut})] }}{(1+z)H(z)} \,,
\label{eq:star-formation redshift distribution}
\end{eqnarray}   
where the normalization constant, $\mathcal{N}_{\rm SFH}$, is determined from $\int dz \, c \, N_{\rm SFH} (z) = 1$, and
\begin{eqnarray}
\dot{\rho}_*(z) = h \dfrac{a + b \, z}{1 + (z/s)^d} \,,
\label{eq:star-formation redshift distribution}
\end{eqnarray}
where $a = 0.017$, $b = 0.13$, $s = 3.3$, $d = 5.3$, and $h = 0.7$.

The integrated optical depth for a given redshift distribution is
\begin{eqnarray}
\bar{\tau}_{\rm const, \, SFH} = \int dz \, \tau(M_L, z)\, N_{\rm const, \, SFH} (z) \,,
\label{eq:integrated optical depth}
\end{eqnarray}
where the integral is over the redshift distribution of the FRBs.  If one observes a large number of FRBs, $N_{\rm FRB}$, then the number of FRBs that will be lensed is
\begin{eqnarray}
N_{\rm lensed\,FRBs} = (1 - e^{-\bar{\tau}_{\rm const., \, SFH}}) \, N_{\rm FRB} \,.
\label{eq:lensed FRBs}
\end{eqnarray}
If none of the FRBs is found to be lensed, then the fraction of dark matter in the form of these dark matter lens can be estimated to be less than $1/ N_{\rm lensed\,FRBs}$.  We assume that $N_{\rm FRB}$ = 10$^4$, a number which can be detected by CHIME\,\cite{2018ApJ...863...48C}.  As will be seen later, even having a sample of $\sim$ 2000 --- 3000 FRBs (with redshift information) will probe new parts of the parameter space for PBHs and exotic compact boson and fermion stars.

The above formalism is valid if the mass distribution of the lens is concentrated in a single mass value: the single mass distribution.  It has been theoretically shown that primordial black holes (acting as the lens) can also have an extended mass distribution\,\cite{Carr:1975qj, Clesse:2015wea, Carr:2017edp, Hawking:1982ga, Hawking:1987bn, Matsuda:2005ez, Berezin:1982ur}.  Constraints on primordial black holes can differ substantially if the underlying mass distribution is extended\,\cite{Carr:2016drx, Carr:2017jsz, Green:2016xgy, Bellomo:2017zsr}.  We will consider the power-law mass distribution of primordial black holes.  In this case, the power law mass distribution is parametrized as\,\cite{Bellomo:2017zsr}
\begin{eqnarray}
\dfrac{d\Phi_{\rm PBH}}{dM} = \dfrac{\mathcal{N}_{\rm PL}}{M^{1-\gamma}} \, \Theta(M_{\rm max} - M) \,\Theta(M - M_{\rm min}),
\label{eq: power law mass distribution}
\end{eqnarray}
where the mass range of the distribution is bordered by the minimum mass, $M_{\rm min}$, and maximum mass, $M_{\rm max}$.  The exponent of the power law is denoted by $\gamma$.   The normalization factor $\mathcal{N}_{PL}$ is
\begin{equation}
\mathcal{N}_{PL} = \left\lbrace
\begin{aligned}
&\frac{\gamma}{-M^\gamma_\mathrm{min} + M^\gamma_\mathrm{max}}, &\gamma\neq0,	\\
&\Bigg[\log \left(\frac{M_\mathrm{max}}{M_\mathrm{min}}\right)\Bigg]^{-1},	&\gamma=0.
\end{aligned}\right.
\label{eq:normPL}
\end{equation}
where the exponent $\gamma$ is determined by the formation epoch of the primordial black hole.  The power law extended mass function is formed in scenarios of cosmic strings and from large density fluctuations.

Due to an extended mass function, the optical depth for lensing (Eq.\,\ref{eq:optical depth eqn 1}) changes and can be written as
\begin{eqnarray}
\tau( z_s) &=& \int dM \int_0 ^{z_s} \, d\chi(z_L) (1 + z_L)^2 n_L \sigma (M, z_L)  \nonumber\\
&\times& \dfrac{d\Phi_{\rm PBH}}{dM} \,.
\label{eq:optical depth extended mass distribution}
\end{eqnarray}
For a power law extended mass distribution, this can be rewritten as
\begin{eqnarray}
&&\tau (M_{\rm min}, M_{\rm max}, z_s) = \int_{M_{\rm min}} ^{M_{\rm max}} dM \, \dfrac{3}{2} f_{\rm DM} \, \Omega_c \nonumber\\
&\times& \int_0 ^{z_s} dz_L \, \dfrac{H_0^2}{c \, H(z_L)} \dfrac{D_L D_{LS}}{D_S} \, (1+z_L)^2  \nonumber\\
&\times& \Big[y_{\rm max}^2(\overline{R}_f) - y_{\rm min}^2(M, \,z_L) \Big]  \dfrac{\mathcal{N}_{\rm PL}}{M^{1-\gamma}} \,.
\label{eq:optical depth power law extended mass distribution}
\end{eqnarray}
This calculated value of the optical depth can be used to calculate the integrated optical depth using Eq.\,\ref{eq:integrated optical depth}, and the number of lensed FRBs (using Eq.\,\ref{eq:lensed FRBs}).

\subsection{Exotic compact objects}
\label{sec:exotic compact objects}
%---------------------------------------

The existence of exotic compact objects has been conjectured for a long time\,\cite{Jetzer:1991jr, Liddle:1993ha, Schunck:2003kk, Liebling:2012fv}, and there has also been a recent surge of interest in this field\,\cite{Chavanis:2011cz, Giudice:2016zpa, Croon:2018ybs, Gresham:2018rqo, Widdicombe:2018oeo, Hardy:2016mns, Maselli:2017vfi, Eby:2015hsq, Braaten:2015eeu, Braaten:2016dlp, Chavanis:2017loo, Eby:2018dat, Chang:2018bgx}.  Very soon after the first discovery of binary black hole merger, it was realized that gravitational waves can also be an efficient way to search for these objects\,\cite{Giudice:2016zpa}.  We will not attempt to systematically go through all the proposals mentioned in the literature.  Instead, we will concentrate on some of the models presented in Ref.\,\cite{Giudice:2016zpa} and indicate the particle physics parameter space which lensing of FRBs can probe.

We will briefly review some of the properties of boson stars and fermion stars closely following Ref.\,\cite{Giudice:2016zpa}.

\subsubsection{Boson stars}

It can be shown via general relativistic calculations that both scalars and vectors can form compact objects\,\cite{Kaup:1968zz, Brito:2015pxa}.  For non-interacting scalars, the star is stabilized due to competition between gravitational attraction and quantum mechanics.  The maximum mass of the boson stars is $M_{\rm max} \approx$ (10$^{-10}$ eV/$m_B$) $M_\odot$\,\cite{Kaup:1968zz}, where $m_B$ is the mass of the boson.  The maximum compactness of the boson star can be calculated to be $C_{\rm max}$ = $(M/ R)_{\rm max}$ $\approx$ 0.08.  For the equilibrium solution, $C \approx$ 0.08 at $M \approx$ 0.85\,$(10^{-10}\, {\rm eV}/m_B)$ M$_\odot$.  Thus, a light boson with mass $10^{-11}$ eV can form a compact star with mass 8.5 M$_\odot$ and radius $\sim$ 158 km.  Such an object is compact enough so that it induces lensing of FRBs --- thus, this technique can act as an efficient probe of the underlying particle physics scenario.

Adding a repulsive self-interaction changes the equilibrium mass - radius relationship.  Assuming a $\lambda |\phi|^4$ interaction (where $\phi$ is the scalar particle and $\lambda$ is the coupling constant), the maximum mass of self-interacting boson star is $M_{\rm max} \approx \, \sqrt{\lambda} \, (100\, {\rm MeV}/m_B)^2 \, \times 10\, M_\odot$ \cite{Colpi:1986ye, Lee:1988av}.  The maximum value of the compactness in this scenario is approximately 0.16.  A further constraint on this scenario comes from self-interaction constraints on dark matter.  Assuming that the self-interaction, $\sigma_{\rm SI}$ in this model is in an astrophysically interesting region (0.1 cm$^2$ g$^{-1} \, \lesssim \sigma_{\rm SI}/ m_B \, \lesssim$ 1 cm$^2$ g$^{-1}$), we obtain the constraint: $(m_B/ {\rm MeV})^{1.5} \lesssim \lambda/10^{-3} \lesssim 3 \times (m_B/ {\rm MeV})^{1.5}$\,\cite{Giudice:2016zpa}.  The range for the self-interaction is motivated by the following observations: as $\sigma_{\rm SI}/ m_B$ becomes less than 0.1 cm$^2$ g$^{-1}$, dark matter interactions are unable to solve the small-scale structure problems in $\Lambda$CDM.  Observational data indicate that $\sigma_{\rm SI}/ m_B \lesssim$ 1 cm$^2$ g$^{-1}$ for a range of velocity scales in astrophysics\,\cite{Tulin:2017ara}.

In order to have a feel for the size of such a compact object, let us define following Ref.\,\cite{Giudice:2016zpa}
\begin{eqnarray}
M_* = \Bigg(\dfrac{M}{1.64 \times 10^6 \, M_\odot} \Bigg)\, \Bigg(\dfrac{m_B}{\rm MeV} \Bigg)^2 \Bigg(\dfrac{4\pi}{\lambda} \Bigg)^{0.5} \,,
\label{eq:Mstar repulsive self-interactions boson star}
\end{eqnarray}
where the maximum value of $M_*$ is approximately 0.22.  In this scenario, the maximum value of the compactness is approximately 0.158.  For $m_B$ = 10 MeV, $\lambda$ = 0.05, and assuming $M_*$ = 0.22, we get that the mass of the compact object is approximately 227 $M_\odot$.  At the maximum value of compactness, this corresponds to a radius of 2130 km.  For perturbativity, we restrict $|\lambda| \lesssim 4\pi$, which implies $m_B \lesssim$ 100 MeV\,\cite{Eby:2015hsq}.

\subsubsection{Fermion stars}

For the discussion of fermion stars, we closely follow the treatment in Refs.\,\cite{Kouvaris:2015rea, Giudice:2016zpa} which considers a model involving self-interacting dark matter and fermionic stars.  It has been shown that if dark matter self-interaction is assumed to solve the small scale structure problems, then the underlying interaction cross-section must be velocity dependent: such a scenario may involve a light mediator or a near threshold S-wave resonance\,\cite{Spergel:1999mh, Tulin:2017ara, Kaplinghat:2015aga, Braaten:2013tza, Laha:2013gva, Laha:2015yoa, Braaten:2018xuw}.

The particle physics model considered in Refs.\,\cite{Kouvaris:2015rea, Giudice:2016zpa} involved a fermionic dark matter candidate, $\chi$, which couples to a vector mediator $V_\mu$ via the interaction $g V_\mu \bar{\chi} \gamma^\mu \chi$.  The equilibrium solution of the mass - radius relation (obtained by solving the Oppenheimer - Volkoff equation) in this case depends on the values of the dark matter mass, $m_F$, mediator mass, $m_V$, and coupling constant, $g$.  The parameter space chosen by the authors in Ref.\,\cite{Giudice:2016zpa} is such that the self-interaction cross section between the two dark matter particles is velocity dependent and obeys the constraint $\sigma/ m_F$ = 0.1 -- 1 cm$^2$\, g$^{-1}$ at velocity scales relevant for dwarf galaxies ($\approx$ 10 km s$^{-1}$).   

We will consider the same parameter space and outline the particle physics parameters which can be probed by FRB lensing. 

For both boson stars and fermion stars, the theory is not yet able to predict the mass spectrum of these compact objects.  As such, a wide variety of masses need to be probed in order to test the underlying physics.  As can be shown by solving the relevant equation, the equilibrium values of the mass and radius of these objects (for a wide range of the particle physics parameters) are such that these can be probed by FRB lensing.  Compared to the lensing constraints from EROS-2\,\cite{Allsman:2000kg} and MACHO\,\cite{Tisserand:2006zx} Collaborations, the lensing of FRBs is sensitive to heavier compact objects ($\gtrsim$ 1 $M_\odot$), and thus these constraints can also be viewed as complementary probes of exotic compact objects.  The constraints from CMB observations which provide the strongest constraint for heavy mass PBHs are not applicable for these exotic compact objects and this underscores the importance of the research of FRB lensing.  The dynamical constraints from the dwarf galaxies\,\cite{Brandt:2016aco, Koushiappas:2017chw} can also probe these exotic compact boson and fermion stars.  These constraints are less reliable when the exotic compact boson and fermion stars make up only a small fraction of the dark matter density or in the presence of the central massive black hole.  The constraint coming from FRB lensing is cleaner and does not suffer from the above mentioned drawbacks.

%-----------------------------------------------------------------------------
\section{Results}
\label{sec:results}
%-----------------------------------------------------------------------------

\begin{figure*}
  \centering
  \includegraphics[width=0.49\textwidth]{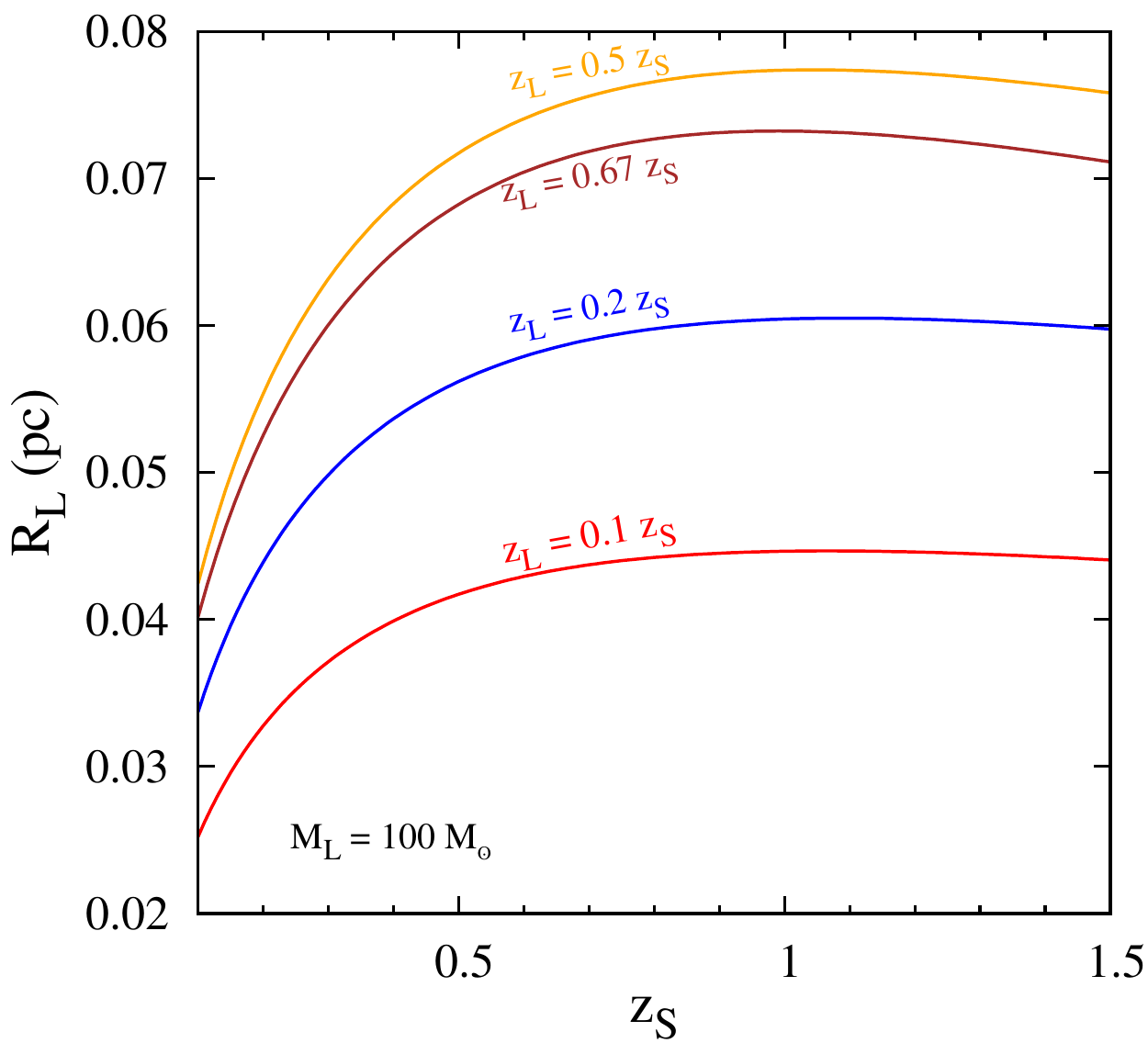}
  \includegraphics[width=0.49\textwidth]{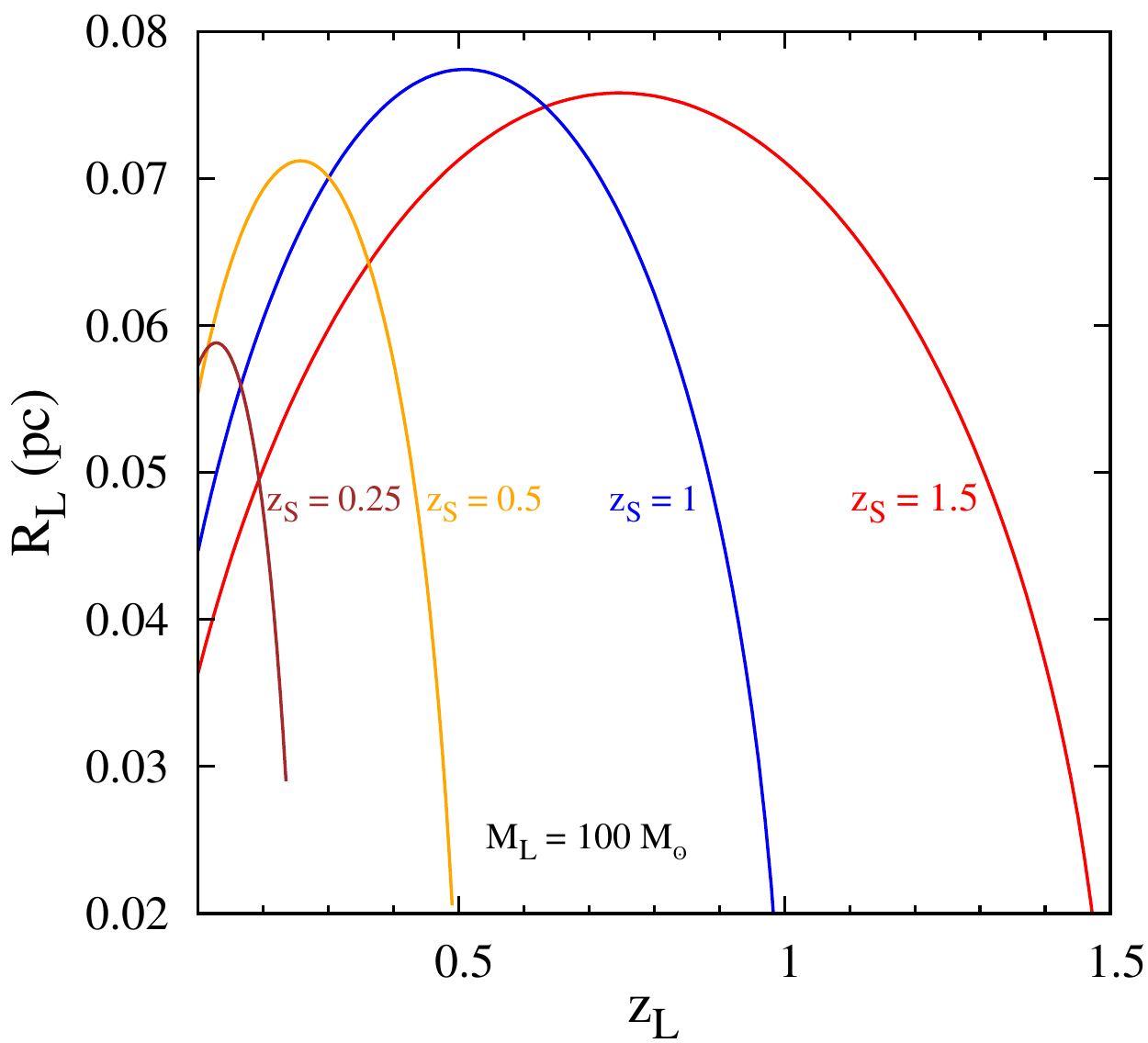}
  \caption{Upper limit on the lens radius (in parsec) in order for the point mass lens approximation to be valid.  We have assumed that the lens mass, $M_L$ = 100 $M_\odot$ for both these panels.  {\bf Left panel:} here we vary the redshift of the source, $z_S$, and fix the the redshift of the lens, $z_L$, to some representative values: $z_L$ = 0.1 $z_S$, $z_L$ = 0.2 $z_S$, $z_L$ = 0.5 $z_S$, and $z_L$ = 0.67 $z_S$.  {\bf Right panel:} here we take some fixed values of the source redshift: $z_S$ = 0.25, 0.5, 1, and 1.5 and vary the lens redshift, $z_L$, from $z_S/50$ to ($z_S$ - 0.01).}
  \label{fig:Upper limit on radius}
\end{figure*}

In this section, we will describe the various results of our calculations.  Assuming various inputs, we will first display the upper limit on the lens radius for our formalism of FRB lensing to be valid.  We will show the constraints on the fraction of dark matter in the form of primordial black holes or exotic boson or fermion stars for the single mass distribution and the extended mass distribution.  Finally, we will demonstrate the particle physics parameter space which FRB lensing is sensitive to assuming bosons or fermions form exotic compact objects.

\subsection{Single mass distribution}

Since we have assumed a point lens in our calculations, we can estimate the upper limit on the size of the lens using Eq.\,\ref{eq:size of lens for point mass approximation} for this assumption to be valid.  The upper limit is proportional to the square root of the mass of the lens, and we assume that $M_L = 100\,M_\odot$ in our plot.  Assuming various choices of $z_S$ and $z_L$, we show this upper limit in Fig.\,\ref{fig:Upper limit on radius}.  In the left panel, we vary $z_S$ from 0.1 to 1.5, and show the upper limit on the lens radius for various positions of $z_L$: $z_L$ = 0.1 $z_S$, $z_L$ = 0.2 $z_S$, $z_L$ = 0.5 $z_S$, and $z_L$ = 0.67 $z_S$.  For each of these choices, the upper limit on the lens radius increases for increasing $z_S$ and then asymptotes out to a constant value.  In the right panel of Fig.\,\ref{fig:Upper limit on radius}, we take $M_L = 100\,M_\odot$, assume some fixed values of $z_S$ = 0.25, 0.5, 1, and 1.5, and vary $z_L$ from $z_S/ 50$ to ($z_S$ - 0.01).  For all the cases we consider, the maximum value of the upper limit on $R_L$ occurs when $z_L \approx 0.5 \, z_S$.  These upper limits on the lens mass are easily satisfied for primordial black holes: the Schwarzschild radius of a 100 $M_\odot$ black hole is $\sim$ 300 km.  For the exotic boson stars and fermion stars that we consider, their masses are typically $\lesssim$ 10 $M_\odot$ and the compactness parameter varies from $\sim$ 0.02 to $\lesssim$ 0.2.  Such an object will have a radius of approximately 741 km to 74 km (corresponding to the two ranges of the compactness parameter that we quoted above).  The values of these radii are much smaller than the upper limits shown in Fig.\,\ref{fig:Upper limit on radius}, thus implying that FRB lensing will constrain the presence of these objects too.

\begin{figure*}
  \centering
  \includegraphics[width=0.49\textwidth]{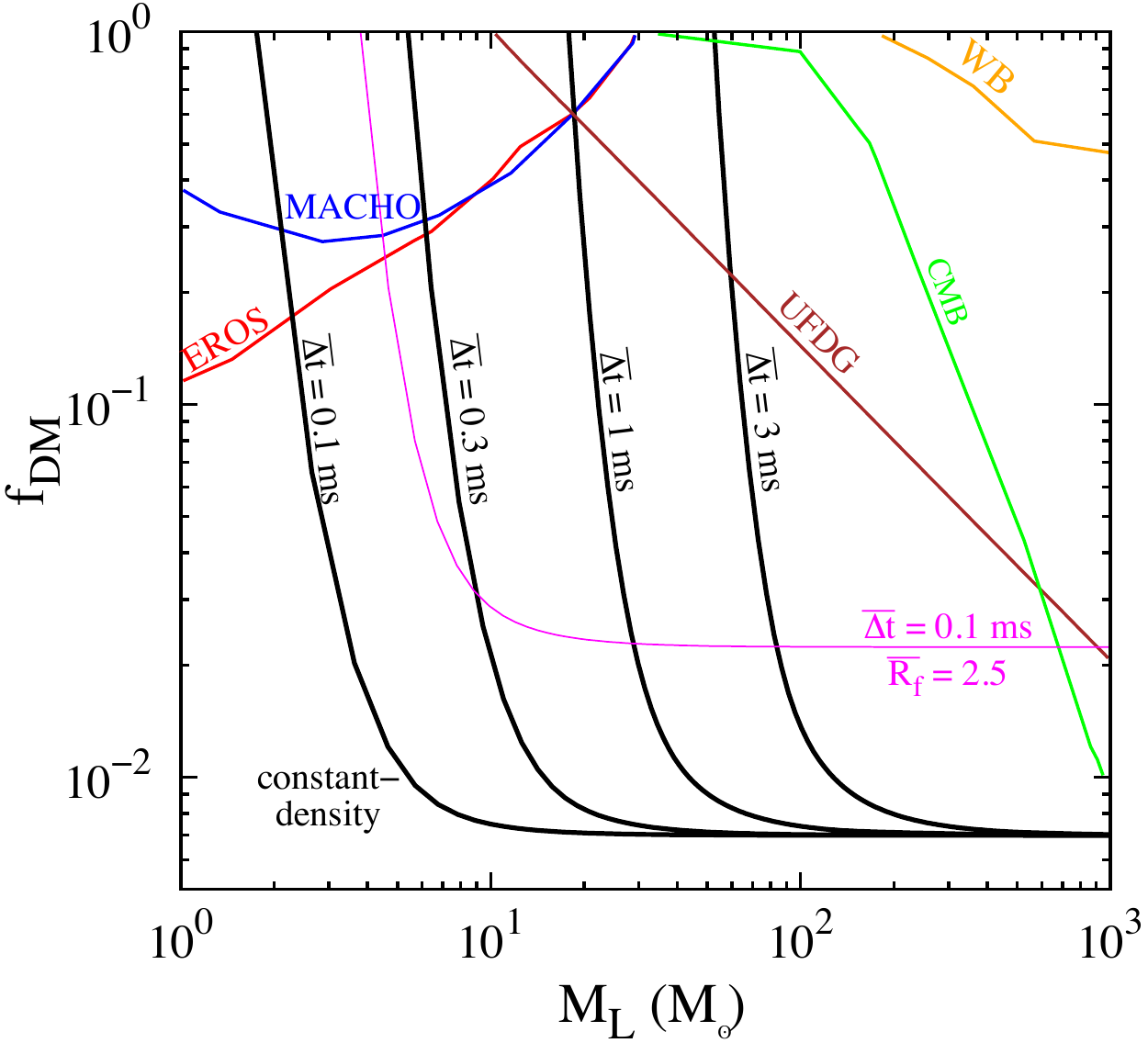}
  \includegraphics[width=0.49\textwidth]{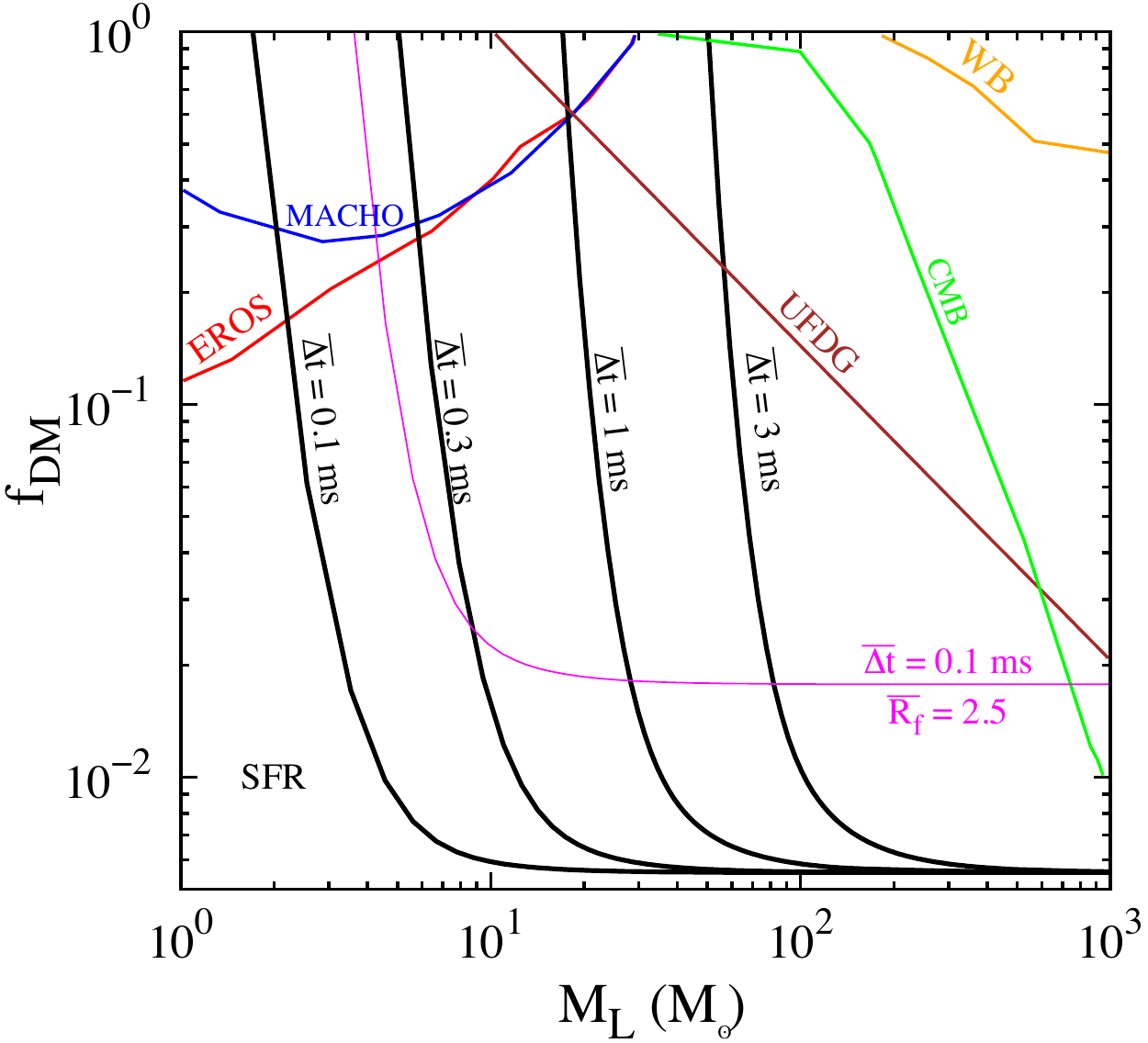}
  \caption{Projected upper limits (in black and magenta solid lines) on the fraction of dark matter, $f_{\rm DM}$, made up of primordial black holes or exotic compact boson/ fermion stars as a function of $M_L$ (in M$_\odot$) that can be achieved via FRB lensing.    The FRB lensing constraints depend on the assumed value of the critical time, $\overline{\Delta t}$ = 0.1 ms, 0.3 ms, 1 ms, and 3 ms.  We use $\overline{R}_f$ = 5 for all our choices of $\overline{\Delta t}$ and additionally use $\overline{R}_f$ = 2.5 for $\overline{\Delta t}$ = 0.1 ms.  The primordial black holes or the exotic compact boson/ fermion stars are assumed to be in a single mass distribution, i.e., the mass distribution at each value is a Dirac delta function.  The left and the right panels assume that the FRBs follow the constant-density redshift distribution and star formation rate (SFR) redshift distribution respectively.  Various other constraints shown in this parameter space come from lensing (MACHO and EROS), kinematic tests (WB and UFDG), and cosmic microwave background (CMB) constraints.  These constraints have been shown following Ref.\,\cite{Bellomo:2017zsr}.  The CMB constraints do not apply to exotic compact boson/ fermion stars.  All the other constraints apply to both primordial black holes and exotic compact boson/ fermion stars.  We assume that the number of FRBs detected to be $10^4$ in order to calculate these projections.}
  \label{fig:fDM_Upper_limit_single_mass_distribution}
\end{figure*}

In Fig.\,\ref{fig:fDM_Upper_limit_single_mass_distribution}, we demonstrate the constraints on dark matter fraction in the form of PBHs or exotic compact boson/ fermion stars for a single mass distribution.  The left and right panels of the figure assume that FRBs follow the constant-density and the star-formation redshift distribution, respectively.  While calculating this result, we assume that $N_{\rm FRB} = 10^4$.  The other constraints shown in the figure are lensing constraints from the EROS and MACHO Collaborations, the dynamical constraints from the wide binary and ultra faint dwarf galaxies, and the constraints from the CMB observations.  The CMB constraints apply only to PBHs, whereas all the other constraints apply to either PBHs or exotic compact fermion/ boson stars.  Given that the temporal width of the bursts varies and the presence of sub-bursts, we estimate that the wide range of $\overline{\Delta t}$ will represent a satisfactory projection into the potential constraints that a future survey like CHIME can obtain.  We have cross-checked that our results for the constant-density redshift distribution matches with that presented in Ref.\,\cite{Munoz:2016tmg}.   

The behaviour of the constraint can be understood by analyzing the equation for the optical depth (eq.\,\ref{eq:optical depth eqn 1}).  In this equation, the only mass dependence comes from the value of $y_{\rm min}$.  The behaviour at high masses can be understood by comparing the values of $y_{\rm min} (M_L, z_L)$ with $y_{\rm max}$.  Assuming $\overline{R}_f = 5$, we get $y_{\rm max} = 0.83$.  The value of $y_{\rm min} (M_L, 0)$ varies from $\sim$ 0.74 to 0.025 for $M_L$ ranging from 10 $M_\odot$ to 300 $M_\odot$ for $\Delta t = 0.3$ ms.  The values at higher redshift are smaller than the corresponding values at lower redshifts but follow the similar trend with respect to the mass dependence.  This behaviour of $y_{\rm min}$ as a function of the mass arises from the fact that for a fixed $z_S$ and $z_L$, the Einstein radius is proportional to the square root of the mass of the lens.  The normalized impact parameter, $y$, is inversely proportional to the Einstein radius, thus explaining the mass dependence of $y_{\rm min}$.  At large lens masses, the values of $y_{min}^2$ for all redshifts are much smaller than $y_{\rm max}^2$, thus explaining the trend that $f_{\rm DM}$ becomes a constant value for appropriate large values of the lens mass.

The threshold of the lens mass at which the lensing constraints are meaningful (i.e., $f_{\rm DM} < 1$) is determined by the value of $M_L$ and $z_L$ at which $y_{\rm min}$ is less than $y_{\rm max}$.  Given our choice of the magnification factor, $\overline{R}_f$, this gives us a constant value of $y_{\rm max}$.  Scanning over the lens mass, the lens mass at which the value of $f_{\rm DM}$ becomes less than or equal to 1 determines the lowest lens mass that can be probed by FRB lensing.  There is a very mild dependence of the threshold lens mass on the FRB redshift distribution, and the major dependence comes from the average widths of the FRB bursts that we have assumed: $\overline{\Delta t}$ = 0.1 ms, 0.3 ms, 1 ms, and 3 ms.  As can be seen from Fig.\,\ref{fig:fDM_Upper_limit_single_mass_distribution}, a smaller value of $\overline{\Delta t}$ allows one to probe lower lens masses.  The choices $\overline{\Delta t}$ = 0.3 ms, 1 ms, and 3 ms are made using the burst durations in \url{frbcat.org} and following Ref.\,\cite{Munoz:2016tmg}.  Given that sub-bursts with much smaller time duration have been observed inside the main burst, we choose $\overline{\Delta t}$ = 0.1 ms and $\overline{R}_f$ = 2.5 or 5 to forecast the limits on $f_{\rm DM}$ that can be achieved if one tries to find the lensing signature of the sub-bursts.  We show the constraints for the $\overline{\Delta t}$ = 0.1 ms using two different values of $\overline{R}_f$ as it will much more challenging to determine the lensing signature of the sub-bursts and using a lower threshold on $\overline{R}_f$ might facilitate the search procedure.  Sub-bursts within the main burst have already been reported in multiple FRBs\,\cite{Champion:2015pmj, Farah:2018buz} and the temporal duration of these sub-bursts is variable.  We still do not know the physical origin of these sub-bursts and the narrowest of these sub-bursts is $\sim$ 30 $\mu$s wide\,\cite{Farah:2018buz}.  If future observations reveal that all or numerous FRBs do indeed contain sub-bursts which are $\mathcal{O}$(10 $\mu$s) wide, then this will permit one to constrain sub-solar mass lens and the corresponding constraints might be competitive or even better that those set by the EROS and MACHO collaboration, and those found by searching for gravitational wave signals\,\cite{Abbott:2018oah}.  Our choice of $\overline{\Delta t}$ = 0.1 ms is not as narrow as the 30 $\mu$s sub-burst, since not all sub-bursts are as narrow as that.

As can be seen from Fig.\,\ref{fig:fDM_Upper_limit_single_mass_distribution}, CHIME will probe sub-percent level of the contribution of PBHs or exotic compact boson/ fermion stars for a wide mass range.  Since the constraints depend linearly on the number of FRBs detected, even a detection of $\sim$ $10^3$ FRBs (with redshift information) will probe unconstrained regions in the $f_{\rm DM}$ - $M_L$ plane.  By using 10$^4$ FRBs, the asymptotic value of $f_{\rm DM}$ that can be probed by this method is $\sim$ 0.7\% and $\sim$ 0.5\% assuming the constant-density redshift distribution and the SFR redshift distribution respectively.  The constraints for the SFR redshift distribution are slightly more stringent as it has a stronger redshift dependence when compared to the constant-density redshift distribution.

\subsection{Extended mass distribution}

\begin{figure*}
  \centering
  \includegraphics[width=0.49\textwidth]{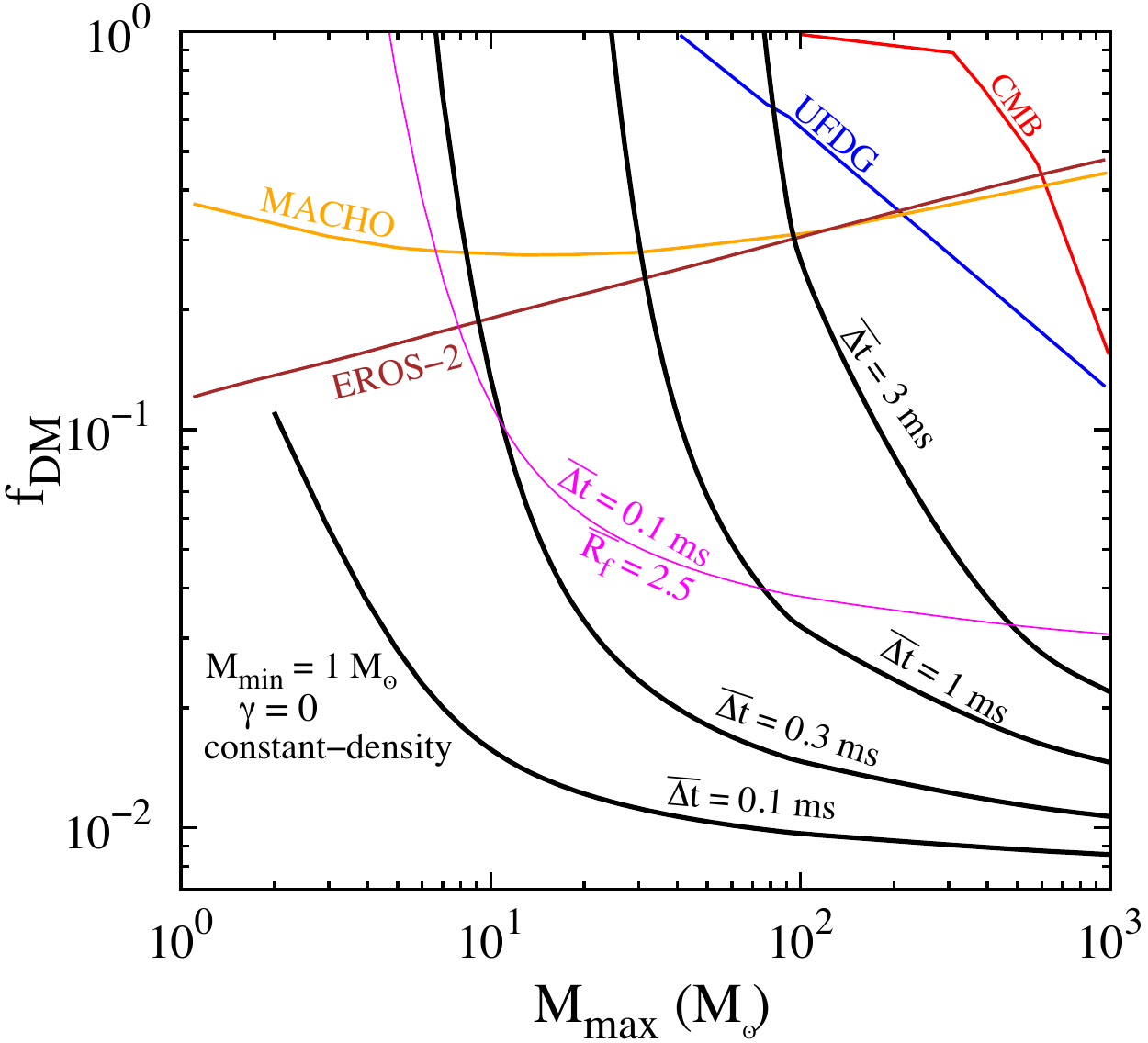}
  \includegraphics[width=0.49\textwidth]{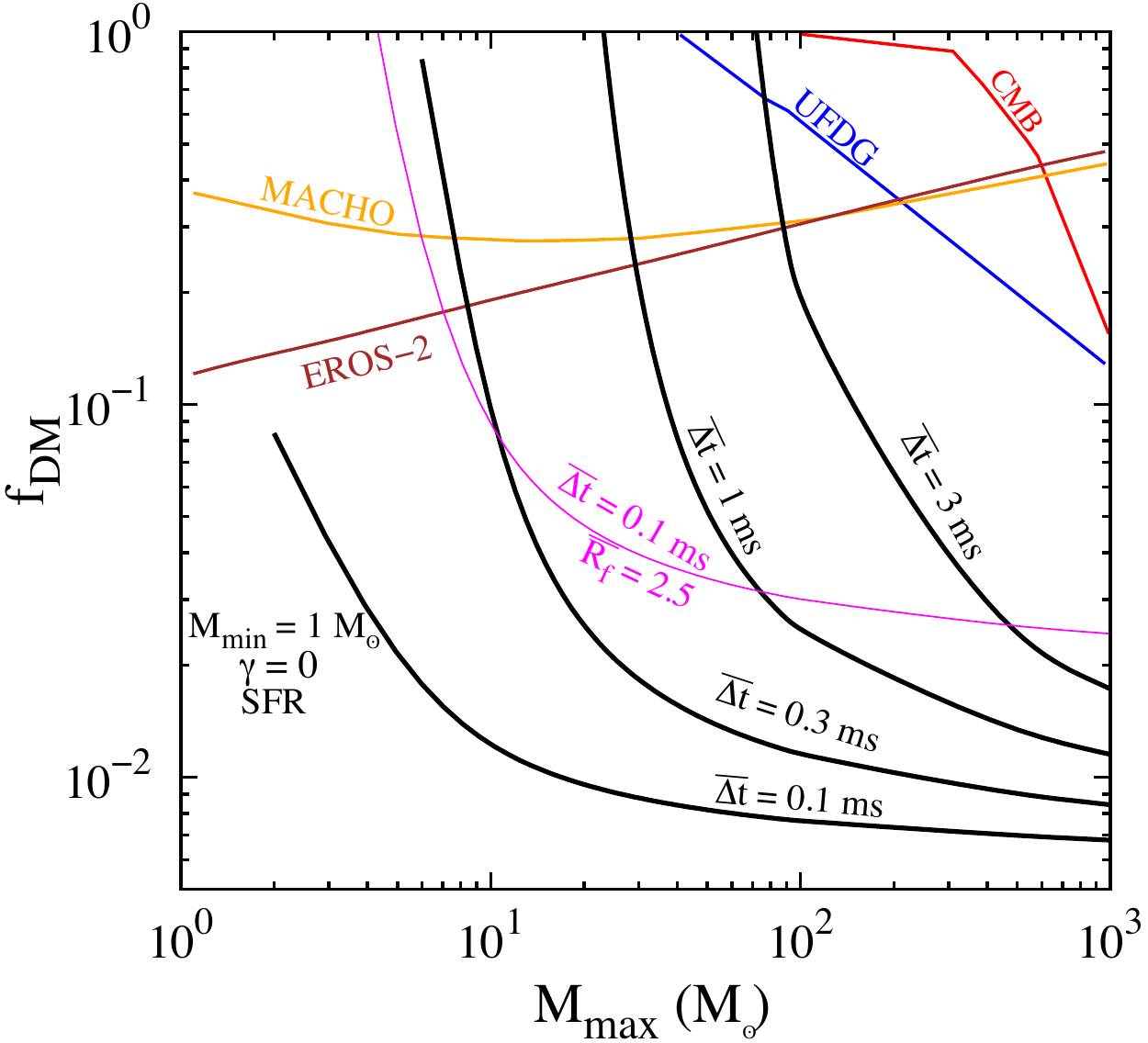}
    \includegraphics[width=0.49\textwidth]{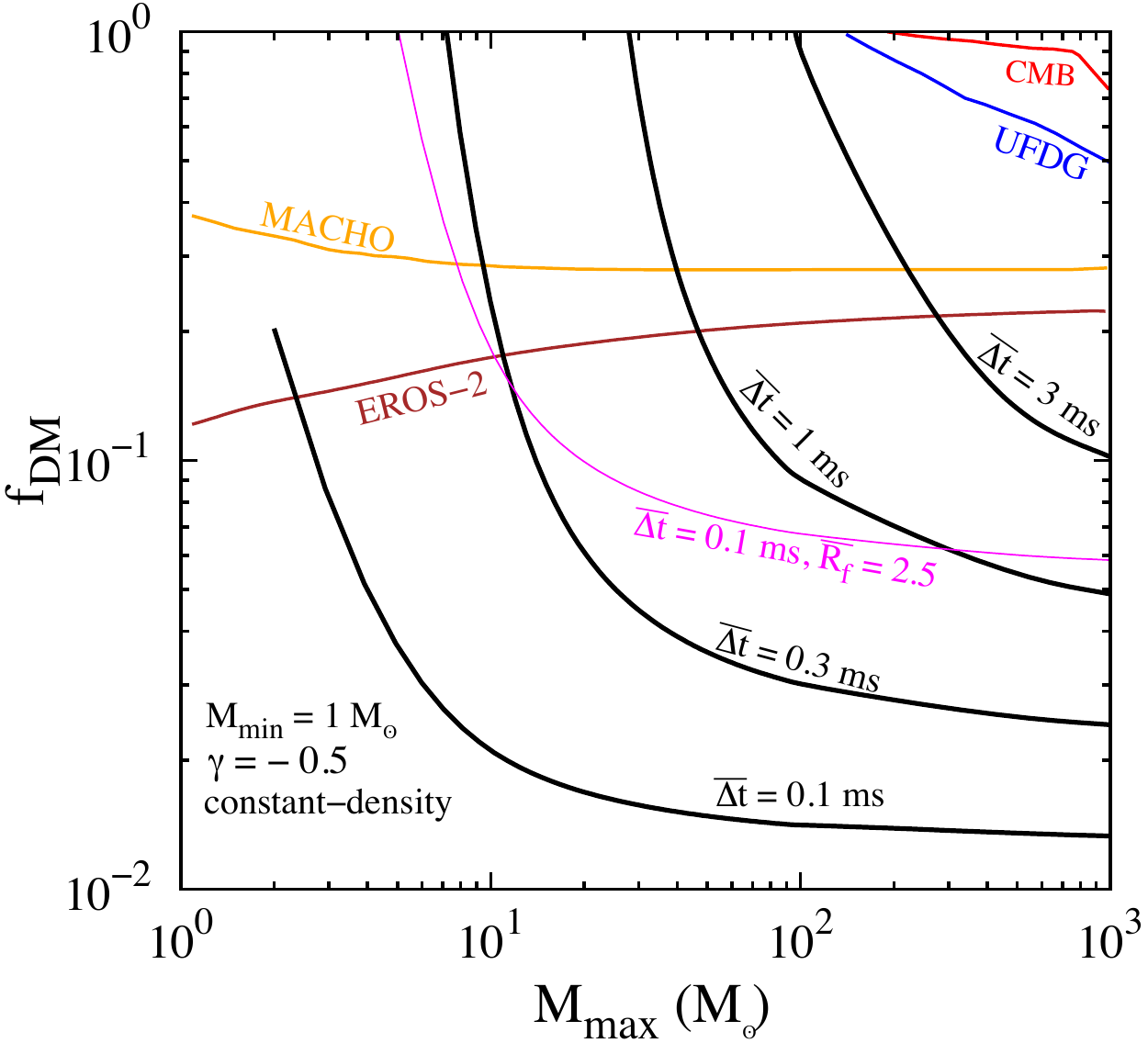}
      \includegraphics[width=0.49\textwidth]{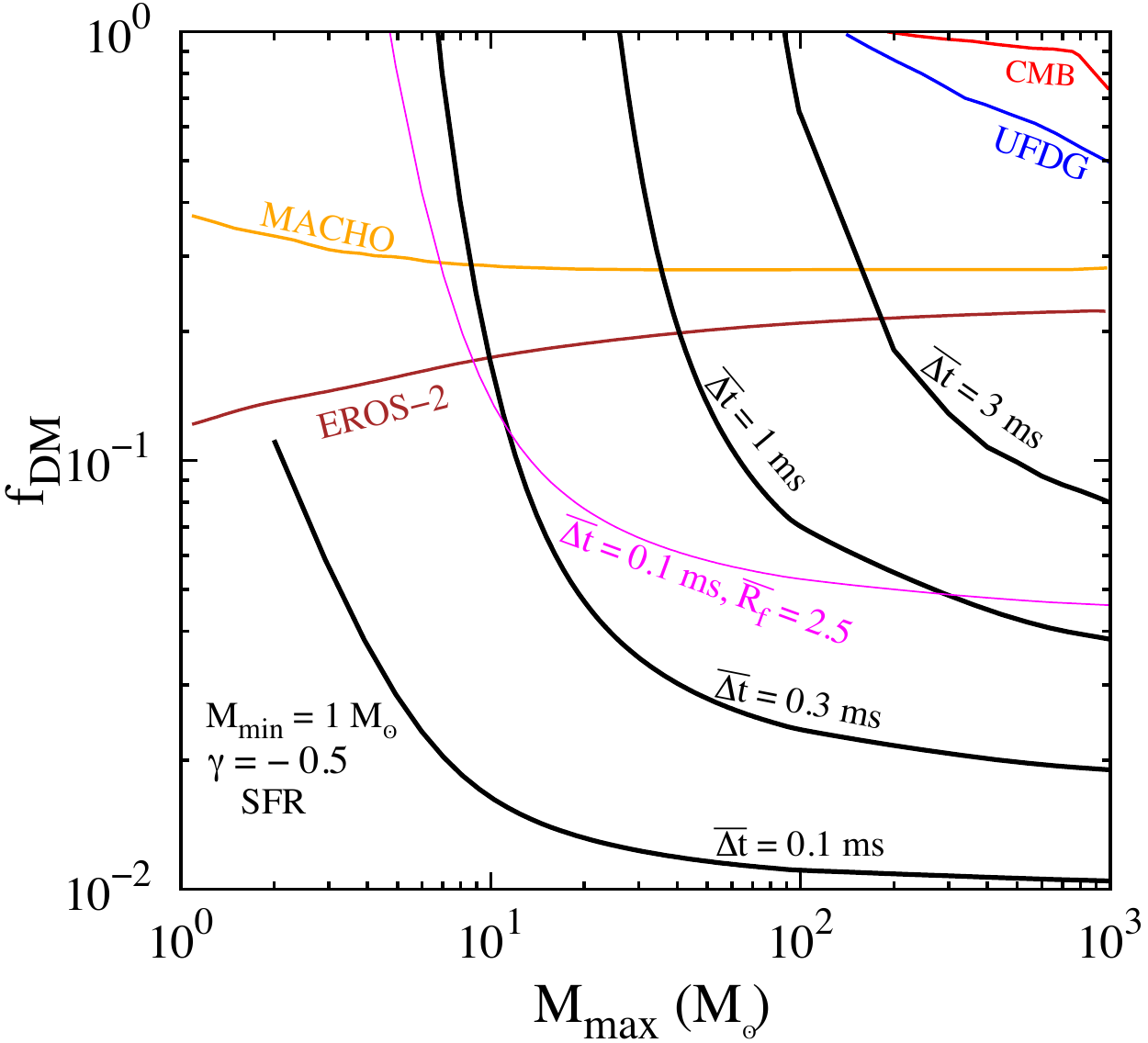}
  \caption{Same as Fig.\,\ref{fig:fDM_Upper_limit_single_mass_distribution} but for extended mass distribution of primordial black holes or exotic compact boson/ fermion stars.  We have assumed a power law mass distribution with a maximum mass of $M_{\rm min}$ = 1 $M_\odot$ in all of these panels.  The exponent of the power law distribution is indicated by $\gamma$.  In these panels, we assume that the power law is bounded by the $M_{\rm min}$ value and the value of the maximum mass, $M_{\rm max}$ indicated in the x-axis.  We have assumed that $M_{\rm max}$ starts from 2 $M_\odot$ and goes up to 10$^3$ $M_\odot$.  Other constraints are plotted following Ref.\,\cite{Bellomo:2017zsr}.}
  \label{fig:fDM_limit_extended_mass_distribution_MMin_1SolarMass}
\end{figure*}

\begin{figure*}
  \centering
  \includegraphics[width=0.49\textwidth]{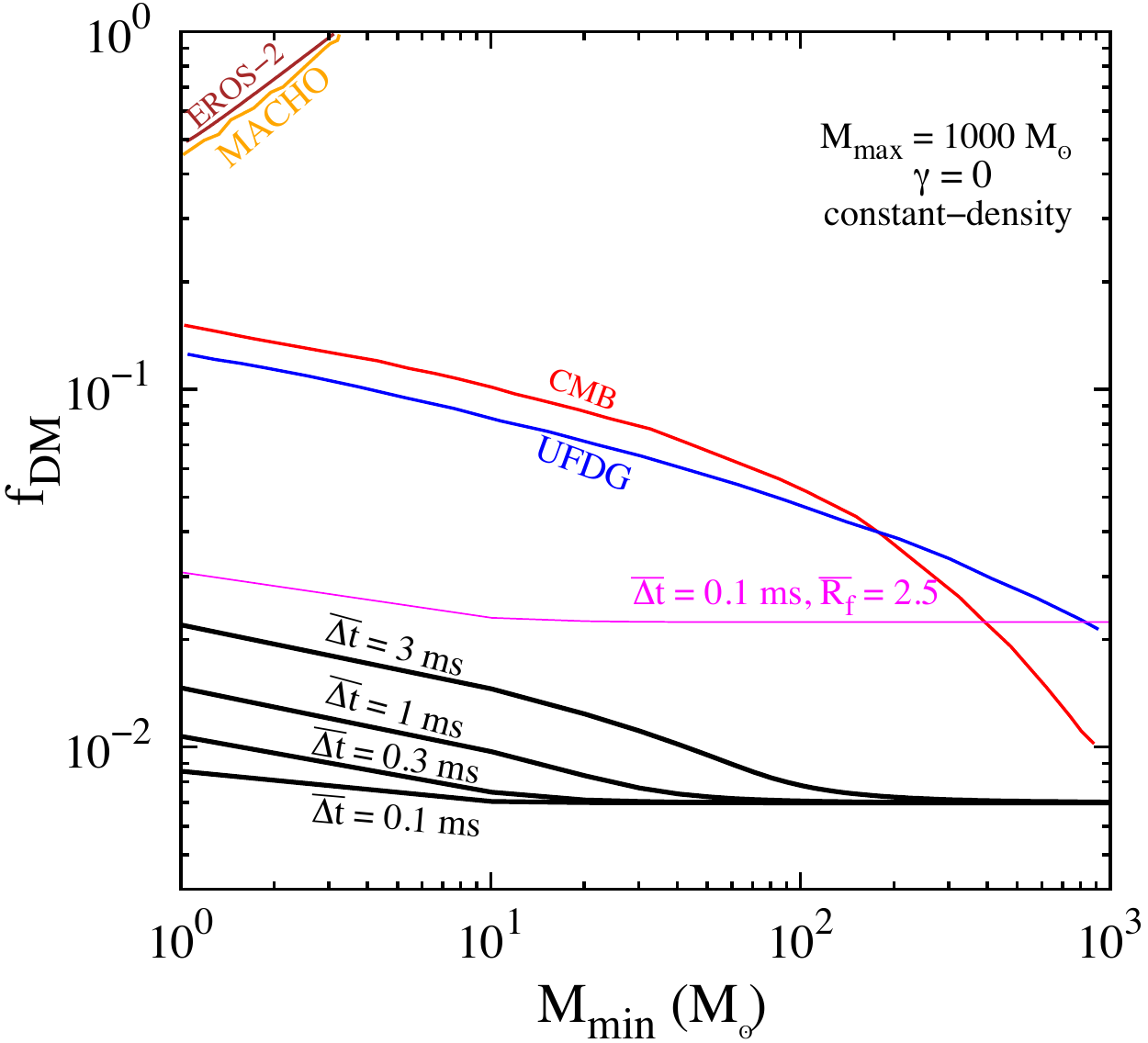}
  \includegraphics[width=0.49\textwidth]{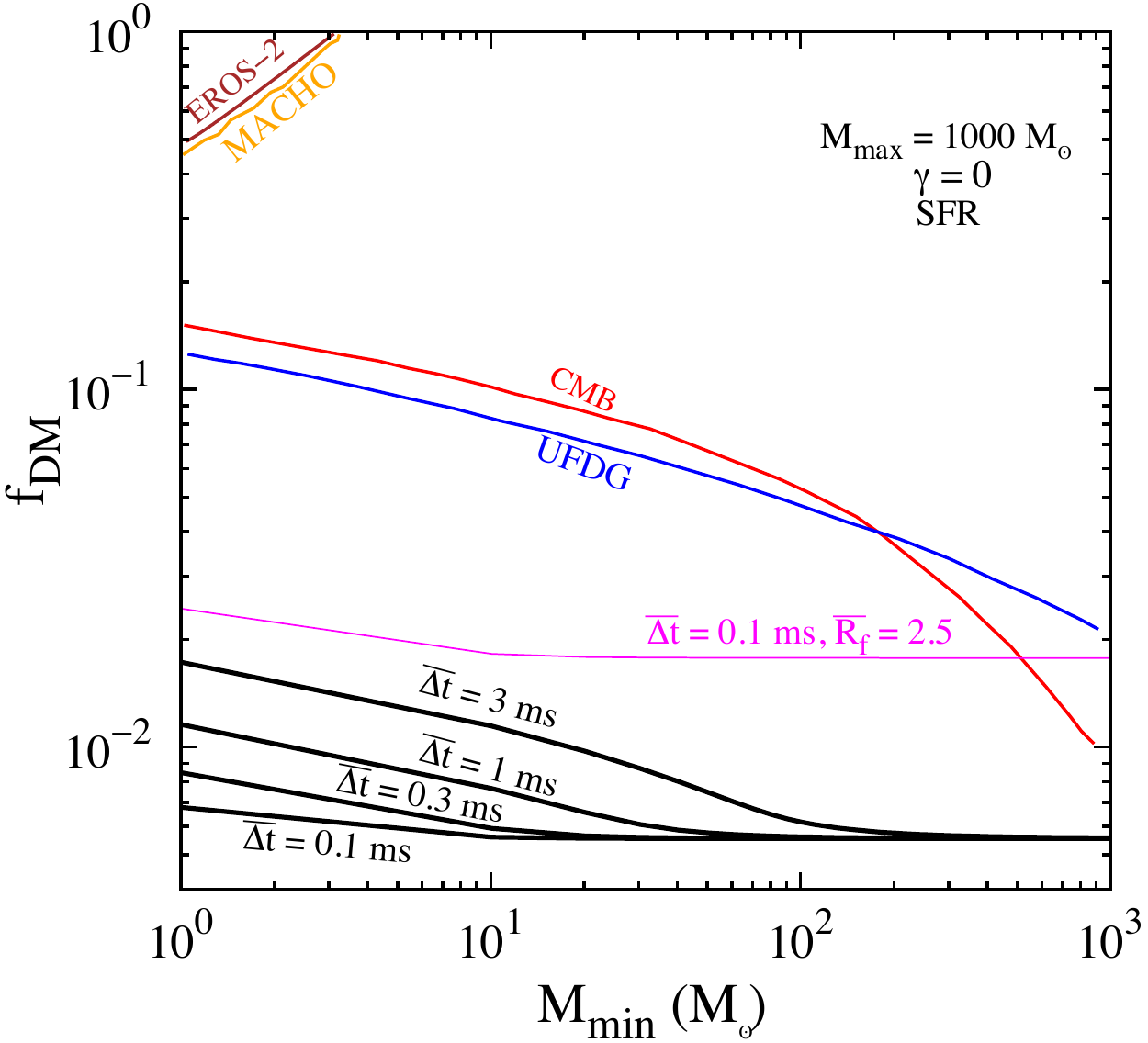}
    \includegraphics[width=0.49\textwidth]{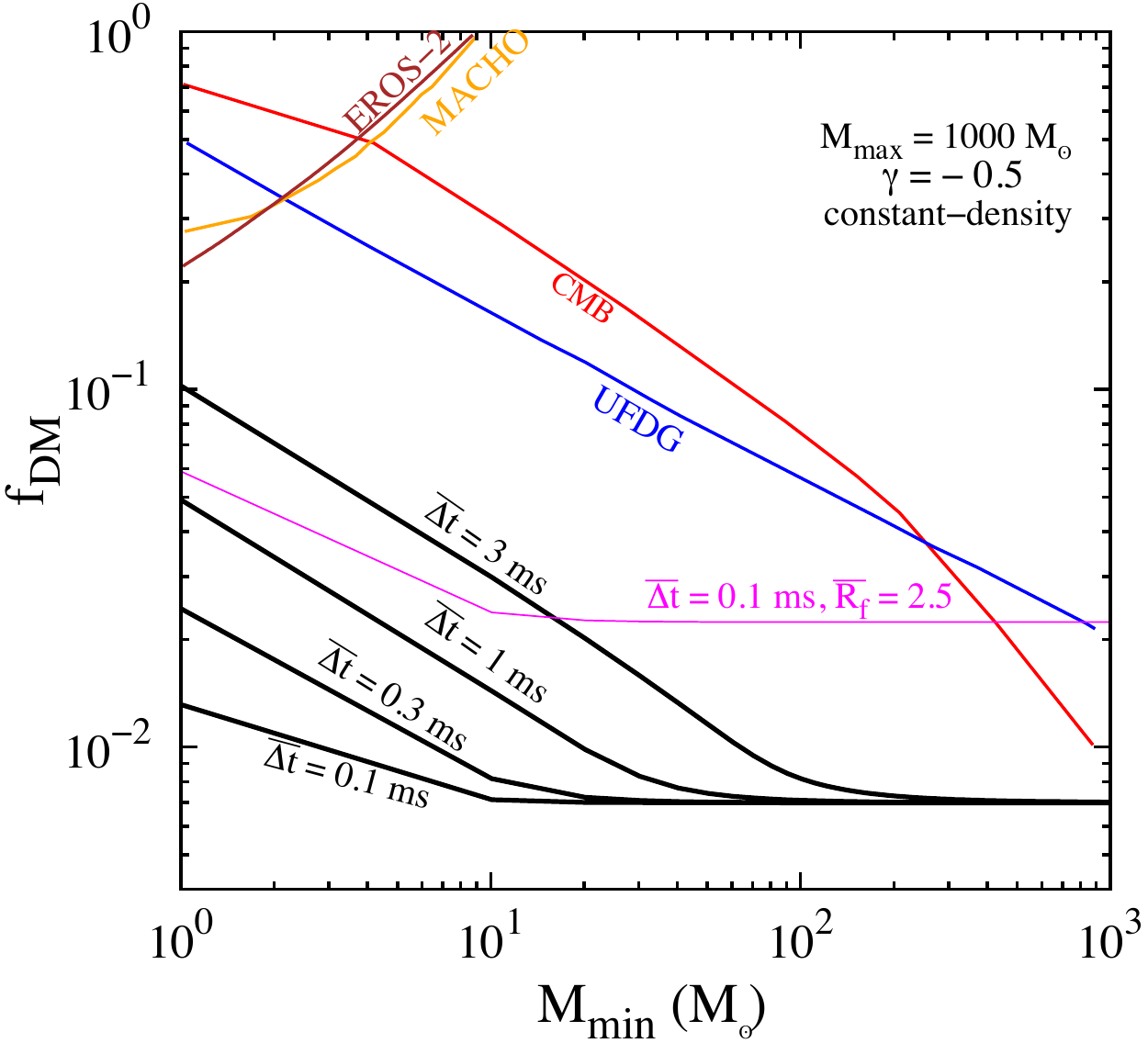}
      \includegraphics[width=0.49\textwidth]{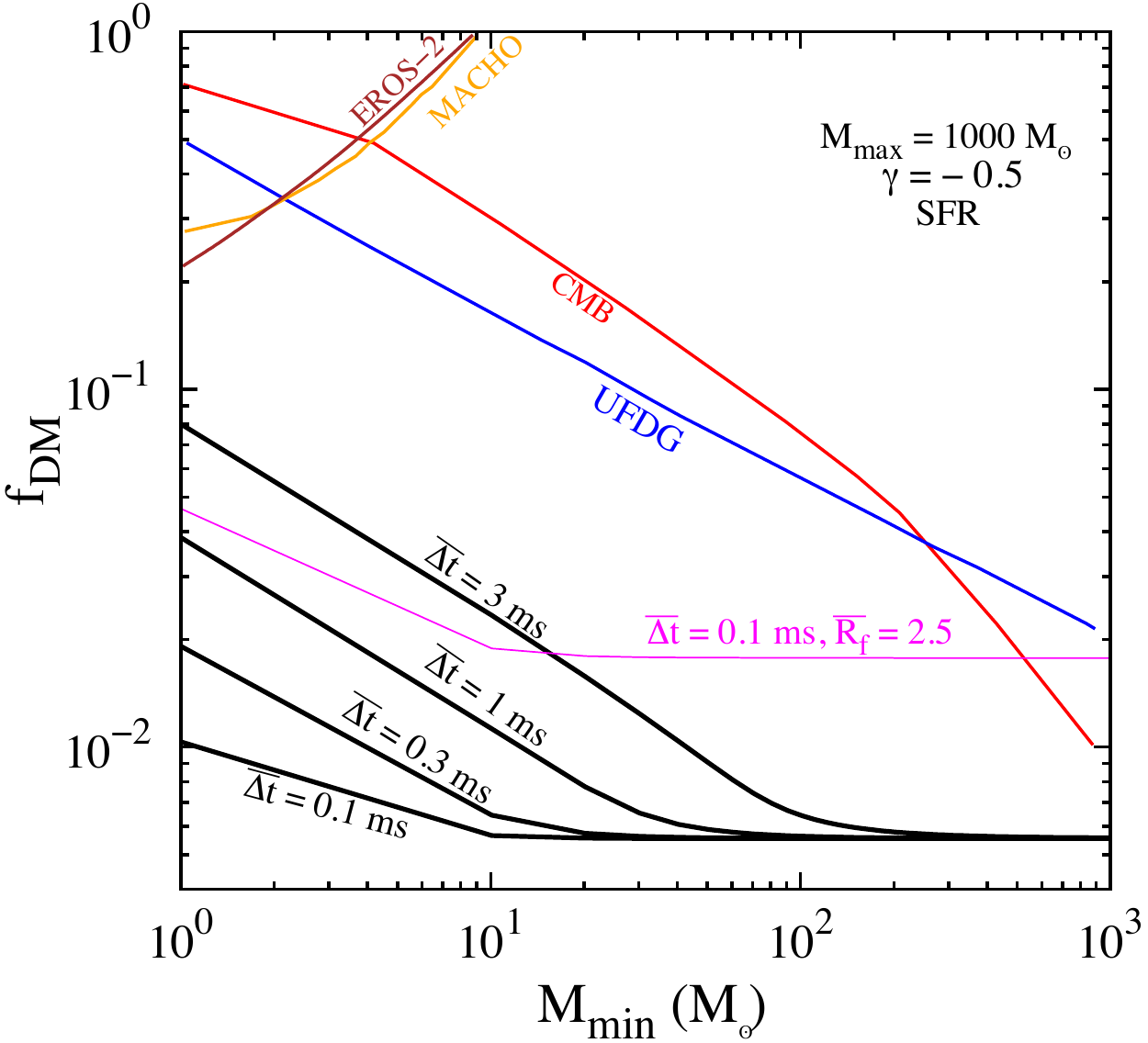}
  \caption{Same as Fig.\,\ref{fig:fDM_limit_extended_mass_distribution_MMin_1SolarMass}.  We have assumed a power law mass distribution with a minimum mass of $M_{\rm min}$ = 1000 $M_\odot$ in all of these panels.  The exponent of the power law distribution is indicated by $\gamma$.  In these panels, we assume that the power law is bounded by the $M_{\rm min}$ value, indicated in the x-axis, and the value of the maximum mass, $M_{\rm max}$.  We have assumed that $M_{\rm min}$ starts from 1 $M_\odot$ and goes up to 999 $M_\odot$.  Other constraints are plotted following Ref.\,\cite{Bellomo:2017zsr}.}
  \label{fig:fDM_limit_extended_mass_distribution_MMax_1000SolarMass}
\end{figure*}

We now derive the projected constraints on $f_{\rm DM}$ assuming that PBHs or exotic compact boson/ fermion stars have an extended mass distribution.  For PBHs, the exact nature of the extended mass distribution is dependent on the underlying inflationary scenario.  Currently, we do not know how to theoretically predict the mass distribution of exotic compact boson/ fermion stars; however, in this sub-section, we will assume that these objects also have an extended mass distribution and calculate the projected constraints.  We choose the power law mass distribution (eqn.\,\ref{eq: power law mass distribution}) as an example to display the projected constraints.  We choose two values of $\gamma$: 0 and -0.5 following Ref.\,\cite{Bellomo:2017zsr}.

The projected constraints are shown in Figs.\,\ref{fig:fDM_limit_extended_mass_distribution_MMin_1SolarMass} and \ref{fig:fDM_limit_extended_mass_distribution_MMax_1000SolarMass}.  In Fig.\,\ref{fig:fDM_limit_extended_mass_distribution_MMin_1SolarMass}, we assume that the minimum value of the extended mass distribution $M_{\rm min}$ = 1 $M_\odot$ and vary the maximum value of the extended mass distribution, $M_{\rm max}$ from 2 $M_\odot$ to 1000 $M_\odot$.  In Fig.\,\ref{fig:fDM_limit_extended_mass_distribution_MMax_1000SolarMass}, we assume that the maximum value of the extended mass distribution, $M_{\rm max}$ = 1000 $M_\odot$ and vary the minimum value of the extended mass distribution, $M_{\rm min}$ from 1 $M_\odot$ to 999 $M_\odot$.  Similar to our choice for the single mass distribution, we take four different values of $\overline{\Delta t}$: 0.1 ms, 0.3 ms, 1 ms, and 3 ms.  We also assume two different redshift distributions for FRBs: constant-density and star-formation rate redshift distributions and assume $N_{\rm FRB}$ = 10$^4$.  Reference\,\cite{Bellomo:2017zsr} has already calculated the current constraints on primordial black holes assuming the power-law extended mass distribution, and we display these constraints in these figures.  

Intuitively, these constraints can be understood by following the equivalent mass formalism introduced in Ref.\,\cite{Bellomo:2017zsr}.  In this formalism, one finds a single lens mass which represents the full effect of the extended mass distribution.    This formalism is valid for any extended mass distributions and Ref.\,\cite{Bellomo:2017zsr} used this formalism to recast the limits on PBHs that had been derived for single mass distribution.  Reference\,\cite{Bellomo:2017zsr} showed that the limits of the dark matter fraction in the form of PBHs can be substantially different for extended mass distributions when compared to single mass distributions.  We find that this equivalent mass formalism is also useful in understanding our results.  

We can calculate the equivalent mass for the extended mass distributions by equating the expressions for the integrated optical depth with appropriate changes,
\begin{eqnarray}
&&\int dz_{\rm mono} \,  \tau (M_{\rm eq}, z_{\rm mono}) N_{\rm evo}(z_{\rm mono}) \nonumber\\
&=& \int dz_{\rm ext} \, \tau (M_{\rm min}, M_{\rm max}, z_{\rm ext}) N_{\rm evo}(z_{\rm ext}) \,,
\label{eq:equivalent mass definition}
\end{eqnarray}
where the equivalent mass is denoted by $M_{\rm eq}$.  The redshifts for the single mass distribution and the extended mass distribution are denoted by $z_{\rm mono}$ and $z_{\rm ext}$, respectively.  The redshift evolution of the FRBs is denoted by the function $N_{\rm evo}$.  After deducing the equivalent mass for a given input of extended mass distribution, we can then read off the $f_{\rm DM}$ for that distribution by using Fig.\,\ref{fig:fDM_Upper_limit_single_mass_distribution}.  Using this method, we can also cross-check the limits on $f_{\rm DM}$ that has been derived following the method outlined in Sec.\,\ref{sec:FRB lensing}.

The shape of the contours can be understood by calculating the equivalent mass for each of the extended mass distributions.  For the extended power-law distribution with $\gamma$ = - 0.5 and $M_{\rm min}$ = 1 $M_\odot$, and the FRBs following the constant-density redshift distribution, the equivalent mass varies from $\sim$ 3.67 $M_\odot$ to $\sim$ 5.4 $M_\odot$, $\sim$ 2.46 $M_\odot$ to $\sim$ 3.4 $M_\odot$, $\sim$ 5.32 $M_\odot$ to $\sim$ 8.1 $M_\odot$, $\sim$ 22.3 $M_\odot$ to $\sim$ 23.8 $M_\odot$, and $\sim$ 52.4 $M_\odot$ to $\sim$ 64.4 $M_\odot$ for $\overline{\Delta t}$ = 0.1 ms ($\overline{R}_f$ = 2.5), 0.1 ms ($\overline{R}_f$ = 5), 0.3 ms, 1 ms, and 3 ms, respectively.  For the same distribution, if the FRBs follow the SFR redshift distribution, then the equivalent masses are in the range of $\sim$ 2.96 $M_\odot$ to $\sim$ 1.96 $M_\odot$, $\sim$ 2.06 $M_\odot$ to $\sim$ 3.6 $M_\odot$, $\sim$ 4.65 $M_\odot$ to $\sim$ 5.8 $M_\odot$, $\sim$ 15 $M_\odot$ to $\sim$ 18.3 $M_\odot$, and  $\sim$ 44.8 $M_\odot$ to $\sim$ 52 $M_\odot$ for $\overline{\Delta t}$ = 0.1 ms ($\overline{R}_f$ = 2.5), 0.1 ms ($\overline{R}_f$ = 5), 0.3 ms, 1 ms, and 3 ms, respectively.  On the other hand, with the same value of $M_{\rm min}$ if the extended mass distribution has $\gamma$ = 0 and the FRBs follow the constant-density redshift distribution, the equivalent mass varies from $\sim$ 3.67 $M_\odot$ to $\sim$ 7.9 $M_\odot$, $\sim$ 2.24 $M_\odot$ to $\sim$ 5.4 $M_\odot$, $\sim$ 5.3 $M_\odot$ to $\sim$ 11.8 $M_\odot$, $\sim$ 17.5 $M_\odot$ to $\sim$ 31.8 $M_\odot$, and $\sim$ 52.3 $M_\odot$ to $\sim$ 83.4 $M_\odot$ for $\overline{\Delta t}$ = 0.1 ms ($\overline{R}_f$ = 2.5), 0.1 ms ($\overline{R}_f$ = 5), 0.3 ms, 1 ms, and 3 ms, respectively.  Assuming the same mass distribution, if the FRBs follow the SFR redshift distribution, the equivalent mass ranges from $\sim$ 2.956 $M_\odot$ to $\sim$ 3.85 $M_\odot$, $\sim$ 1.96 $M_\odot$ to $\sim$ 2.13 $M_\odot$, $\sim$ 4.65 $M_\odot$ to $\sim$ 6.2 $M_\odot$, $\sim$ 14.95 $M_\odot$ to $\sim$ 20 $M_\odot$, and $\sim$ 44.9 $M_\odot$ to $\sim$ 58.6 $M_\odot$ for $\overline{\Delta t}$ = 0.1 ms ($\overline{R}_f$ = 2.5), 0.1 ms ($\overline{R}_f$ = 5), 0.3 ms, 1 ms, and 3 ms, respectively.  Comparing these equivalent mass values with the corresponding upper limits on $f_{\rm DM}$ from Fig.\,\ref{fig:fDM_Upper_limit_single_mass_distribution} can explain our projected limits in Fig.\,\ref{fig:fDM_limit_extended_mass_distribution_MMin_1SolarMass}.

We see that the projected limits on $f_{\rm DM}$ due to FRB lensing will be much stronger than other constraints for both the cases that we consider in Figs.\,\ref{fig:fDM_limit_extended_mass_distribution_MMin_1SolarMass} and \ref{fig:fDM_limit_extended_mass_distribution_MMax_1000SolarMass}.  As in the case of the single mass distribution, the best limits are achieved when $\overline{\Delta t}$ = 0.1 ms.  We assume $N_{\rm FRB}$ = 10$^4$ with redshift information for all these plots.  From the upper panel in Fig.\,\ref{fig:fDM_limit_extended_mass_distribution_MMin_1SolarMass}, we see that even detecting $\sim$ 3000 FRBs with redshift information will probe new parts of the parameter space for the extended power-law distribution with $\gamma$ = - 0.5 and $M_{\rm min}$ = 1 $M_\odot$.  However, for the case of $\gamma$ = 0 and $M_{\rm min}$ = 1 $M_\odot$, one requires 10$^4$ FRBs with redshift information to probe new parts of the parameter space assuming $\overline{\Delta t}$ = 3 ms.  Assuming a smaller $\overline{\Delta t}$ will constrain new parameter space with smaller number of FRB detection.

The equivalent mass values if the extended mass distribution follows a power law with $M_{\rm max}$ = 1000 $M_\odot$ is substantially different from the other case mentioned above.  If the FRBs follow the constant-density redshift distribution and the exponent of the power law is $\gamma$ = - 0.5, the equivalent mass ranges from $\sim$ 5.35 $M_\odot$ to $\sim$ 875 $M_\odot$, $\sim$ 3.5 $M_\odot$ to $\sim$ 800.5 $M_\odot$, $\sim$ 8.2 $M_\odot$ to $\sim$ 295 $M_\odot$, $\sim$ 23.6 $M_\odot$ to $\sim$ 230 $M_\odot$, and $\sim$ 64 $M_\odot$ to $\sim$ 877 $M_\odot$ for $\overline{\Delta t}$ = 0.1 ms ($\overline{R}_f$ = 2.5), 0.1 ms ($\overline{R}_f$ = 5), 0.3 ms, 1 ms, and 3 ms, respectively.  For the same redshift distribution, if the exponent of the power law is $\gamma$ = 0, the equivalent mass is between $\sim$ 8 $M_\odot$ to $\sim$ 865.7 $M_\odot$, $\sim$ 5.7 $M_\odot$ to $\sim$ 855.7 $M_\odot$, $\sim$ 11.8 $M_\odot$ to $\sim$ 989 $M_\odot$, $\sim$ 32 $M_\odot$ to $\sim$ 990.7 $M_\odot$, and $\sim$ 85 $M_\odot$ to $\sim$ 999 $M_\odot$ for $\overline{\Delta t}$ = 0.1 ms ($\overline{R}_f$ = 2.5), 0.1 ms ($\overline{R}_f$ = 5), 0.3 ms, 1 ms, and 3 ms, respectively.  If the FRBs follow SFR redshift distribution and the exponent of the power law is $\gamma$ =  - 0.5, then the equivalent mass varies from $\sim$ 5.2 $M_\odot$ to $\sim$ 809.4 $M_\odot$, $\sim$ 3.6 $M_\odot$ to $\sim$ 689 $M_\odot$, $\sim$ 8 $M_\odot$ to $\sim$ 954 $M_\odot$, $\sim$ 23 $M_\odot$ to $\sim$ 994 $M_\odot$, and $\sim$ 62.5 $M_\odot$ to $\sim$ 999 $M_\odot$ for $\overline{\Delta t}$ = 0.1 ms ($\overline{R}_f$ = 2.5), 0.1 ms ($\overline{R}_f$ = 5), 0.3 ms, 1 ms, and 3 ms, respectively.  If the FRBs follow SFR redshift distribution and the exponent of the power law is $\gamma$ = 0, then the equivalent mass varies from $\sim$ 7.8 $M_\odot$ to $\sim$ 830 $M_\odot$, $\sim$ 5.5 $M_\odot$ to $\sim$ 990 $M_\odot$, $\sim$ 11.6 $M_\odot$ to $\sim$ 999 $M_\odot$, $\sim$ 31.2 $M_\odot$ to $\sim$ 998.2 $M_\odot$, and $\sim$ 81.2 $M_\odot$ to $\sim$ 998.2 $M_\odot$ for $\overline{\Delta t}$ = 0.1 ms ($\overline{R}_f$ = 2.5), 0.1 ms ($\overline{R}_f$ = 5), 0.3 ms, 1 ms, and 3 ms, respectively. 

Comparing these equivalent mass values with the corresponding upper limits on $f_{\rm DM}$ from Fig.\,\ref{fig:fDM_Upper_limit_single_mass_distribution} can explain our projected limits in Fig.\,\ref{fig:fDM_limit_extended_mass_distribution_MMax_1000SolarMass}.  We again see that the projected limits from FRB lensing will be much better than other constraints on PBHs or exotic compact boson/ fermion stars.  We find that the shape of the projected constraints is quite different in Fig.\,\ref{fig:fDM_limit_extended_mass_distribution_MMax_1000SolarMass} compared to Fig.\,\ref{fig:fDM_limit_extended_mass_distribution_MMin_1SolarMass}.  For example, the asymptotic value of the upper limit on $f_{\rm DM}$ is observed in Fig.\,\ref{fig:fDM_limit_extended_mass_distribution_MMax_1000SolarMass} but not in Fig.\,\ref{fig:fDM_limit_extended_mass_distribution_MMin_1SolarMass}.  This is easily explained by the large values of the equivalent masses that we find for the extended mass distribution considered in Fig.\,\ref{fig:fDM_limit_extended_mass_distribution_MMax_1000SolarMass}.
  
\subsection{Probe of exotic compact fermion/~boson stars}

\begin{figure*}
  \centering
  \includegraphics[width=0.328\textwidth]{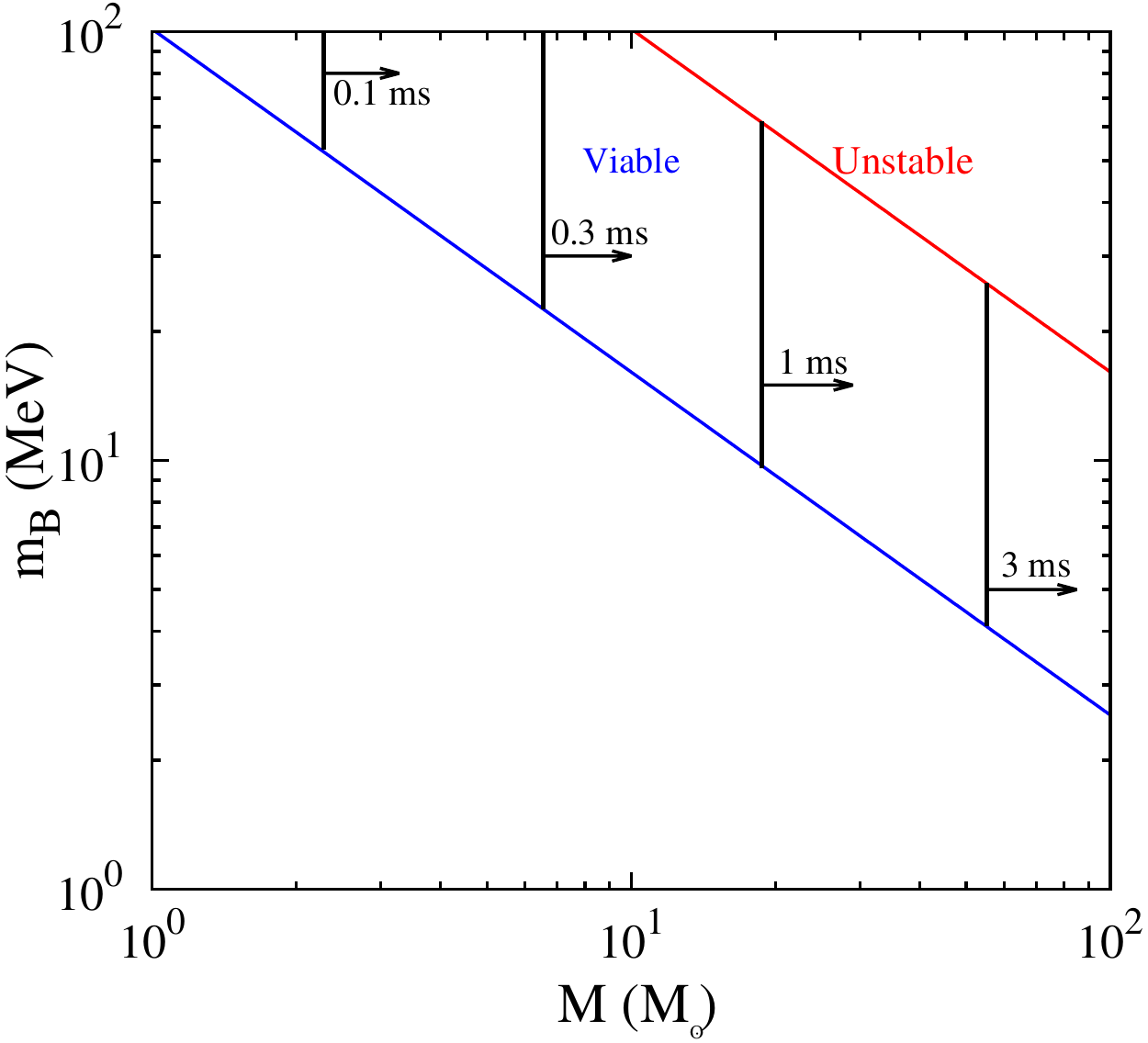}
  \includegraphics[width=0.328\textwidth]{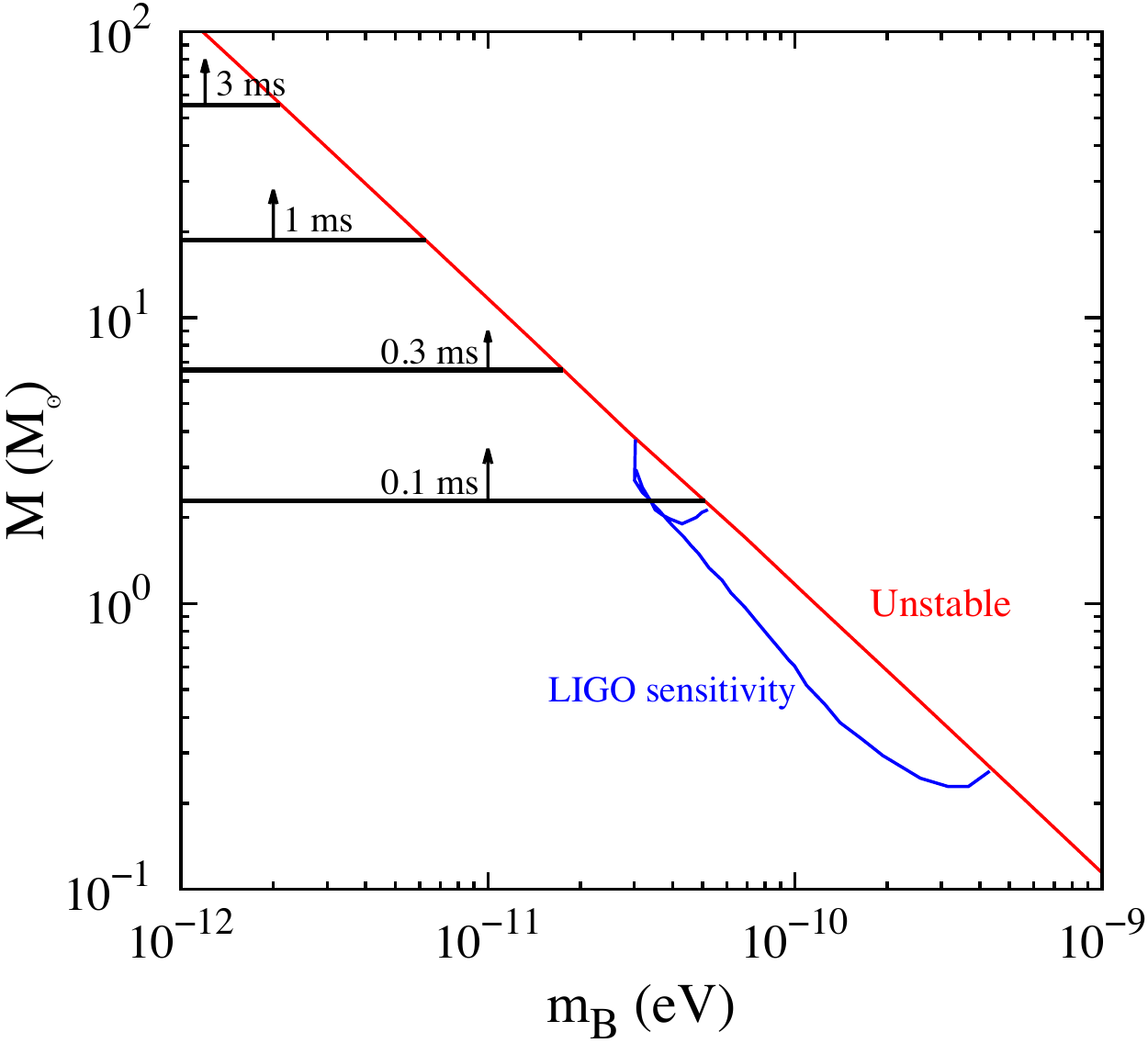}
    \includegraphics[width=0.328\textwidth]{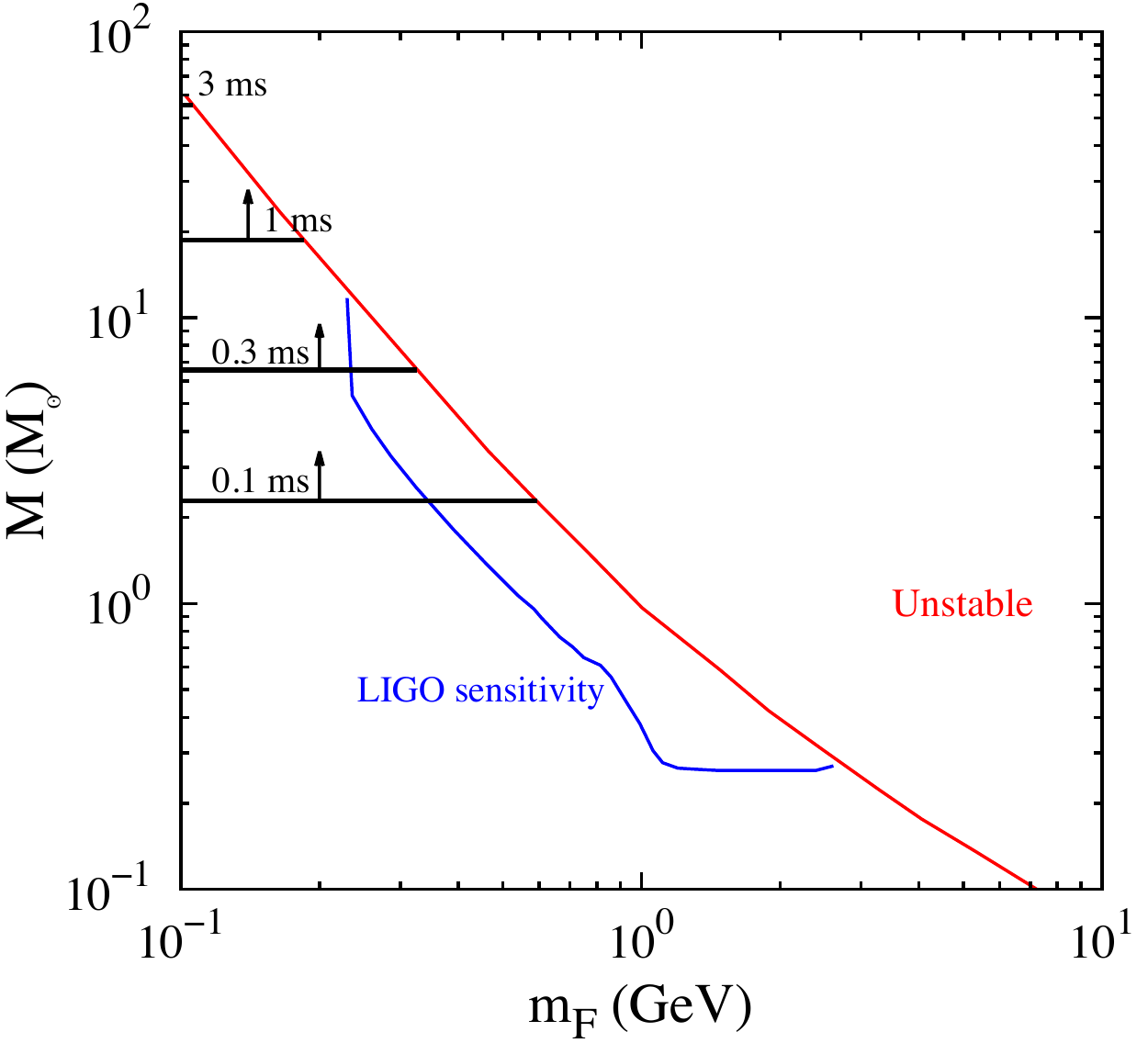}
  \caption{Constraints on various particle physics models which can be derived from FRB lensing.  The parts of the parameter space which do not produce a stable solution for a bound object is marked as ``unstable".  {\bf Left panel:} the parameter space for self-interacting bosons.  The region between the red line and the blue line can produce a boson star and also solve the small-scale structure problems of $\Lambda$CDM.  The region to the right of the black horizontal lines can be probed by FRB lensing assuming various characteristic time delays.  {\bf Middle panel:}  the parameter space for free bosons.  The region bounded by the blue line and the red line can be probed by LIGO observations.  The region above the black horizontal lines can be probed by FRB lensing.  {\bf Right panel:}  the parameter space for self-interacting fermions.  The region that can be probed by LIGO observations is bounded by the blue line and the red line.  The region above the black horizontal lines can be probed by FRB lensing.  In this case there is a tiny amount of parameter space which can be probed by FRB lensing if $\overline{\Delta t}$ = 3 ms (shown by the topmost horizontal line).}
  \label{fig:Particle physics}
\end{figure*}

The formation of exotic compact fermion/ boson stars has been predicted in many beyond the Standard Model particle physics models.  In Fig.\,\ref{fig:Particle physics}, we display the parts of the particle physics parameter space which will be probed by FRB lensing.  We concentrate on the models studied in Ref.\,\cite{Giudice:2016zpa}, but note that there are many other particle physics models which are also predicted to form these exotic compact fermion/ boson stars and FRB lensing can probe a substantial number of these models.  Given the large values of the upper limit on the lens radius as can be seen from Fig.\,\ref{fig:Upper limit on radius}, it is not surprising to find the wide ranges of particle physics parameters which FRB lensing can probe.

In the left panel of Fig.\,\ref{fig:Particle physics}, we display the scalar boson mass v/s the lens mass and the parameter space which can be probed by FRB lensing.  The self-coupling constant, $\lambda$, is varied between $10^{-3} (m_B/{\rm MeV})^{1.5} \leq \lambda \leq 3 \times 10^{-3} (m_B/{\rm MeV})^{1.5}$ in the region between the red line and the blue line.  The region to the right of the red line is excluded because the dimensionless mass, $M_*$, is greater than its maximum possible value 0.22.  We restrict the range of the boson mass, $m_B$, such that the self-coupling is perturbative.  We find that the range of the parameter space displayed in this figure can be probed by FRB lensing, although the exact part of the particle physics parameter space can only be determined after one detects such a lens.  The region to the right of the vertical lines are the parts of the parameter space that can be probed by FRB lensing.  The vertical lines are plotted for the values of $\overline{\Delta t}$ and $\overline{R}_f$ = 5, which have been studied above.  These lines correspond to the lens mass at which $f_{\rm DM}$ = 0.5.  Some of the low lens mass region in this model is already probed by the results from the MACHO and the EROS Collaborations and this figure along with our previous results show how different searches for exotic compact fermion/ boson stars can probe novel particle physics models.

The middle panel of Fig.\,\ref{fig:Particle physics} shows the particle physics parameter space for bosons with no self-interactions.  In this case, the mass of the relevant bosons is much smaller than the electron and searching for exotic compact objects is one of the main avenues of research.  In this plot, the region to the right of the red line is excluded as the compactness is greater than its theoretical maximum value.  The region bounded by the blue line and the red line can be probed by LIGO assuming that the lens is at a luminosity distance less than 450 Mpc (smaller region) and 100 Mpc (larger region)\,\cite{Giudice:2016zpa}.  We show the region that can be probed by FRB lensing via the horizontal solid lines.

The right panel in Fig.\,\ref{fig:Particle physics} shows the parameter space for fermion stars which are formed by clustering of exotic fermions with self-interactions among themselves.  For this figure, we assume $\alpha \equiv g^2/4\pi$ = $10^{-2}$ and the mediator mass, $m_V$, is in the range [$10^{-2} \, {\rm GeV}, 10^{-1} \, {\rm GeV}$].  As has been demonstrated earlier in the literature\,\cite{Tulin:2013teo}, such a parameter range can induce self-interaction between the fermions which can solve the small-scale structure problems in $\Lambda$CDM.  The region that can be probed by FRB lensing is shown by horizontal lines and the region bounded by the blue and the red line shows the LIGO sensitivity region\,\cite{Giudice:2016zpa}.  The region to the right of the red line produces unstable solution and is not considered.  These show the complementarity between the gravitational wave observations and FRB lensing in constraining these exotic compact objects.

%-----------------------------------------------------------------------------
\section{Conclusions}
\label{sec:conclusions}
%-----------------------------------------------------------------------------

The era of gravitational wave observations has started via the pioneering observations of the LIGO - Virgo collaboration.  The initial observations have already shed light on some major astrophysics questions.  It has already been realized that gravitational waves can probe beyond the Standard Model physics questions (see, for e.g., Refs.\,\cite{Kopp:2018jom, Breitbach:2018ddu}).  It has been conjectured that the observed binary black hole mergers are primordial in nature.  Such an observation carries profound implications, and thus numerous studies have been designed to study and distinguish primordial and astrophysical black holes.

Fast radio bursts are one of the major new discoveries in astrophysics and the origin of these $\sim \mathcal{O}$(ms) duration bursts of radio photons is unknown.  With the data from $\sim$ $\mathcal{O}$(100) FRBs that have been discovered till date, it is estimated that these mysterious astrophysical transients are numerous and are probably cosmological in origin.  The sharp feature of the radio burst and its cosmological origin has already prompted numerous studies in the literature which use FRBs as fundamental probes.  

One of the cleanest tests of primordial black holes contributing to the dark matter density of the Universe was proposed in Ref.\,\cite{Munoz:2016tmg}.  In this case, the lensing of FRBs by primordial black holes will produce multiple images of the burst which can be tested via upcoming radio telescopes.  Assuming that the time delay between the two images is greater than the burst time, one can cleanly probe the presence of a compact object near the line of sight.  Assuming a single mass distribution of primordial black holes and a constant-density redshift distribution of FRBs, it has already been shown that this technique can probe new parts of the parameter space and can detect primordial black holes even if they make up a sub-percent level of the dark matter density\,\cite{Munoz:2016tmg}.

We extend this result by explicitly displaying the constraints on dark matter fraction which FRB lensing can probe if primordial black holes (acting as the lens) have a single mass distribution and FRBs follow the star-formation rate redshift distribution (Fig.\,\ref{fig:fDM_Upper_limit_single_mass_distribution} right panel).  We find that even in this case, this technique will produce the leading constraints in a wide range of the lens masses.  The threshold lens mass which this technique can probe will depend on the characteristic time delay, $\overline{\Delta t}$ that we employ in our calculations.  Besides using the standard values of $\overline{\Delta t}$ that has been used in the literature, we also use $\overline{\Delta t}$ = 0.1 ms to simulate the effect of mini-bursts within the main burst.  In the latter case, one can try to detect the lensed image of these mini-bursts.  An advantage of trying to detect the lensing signatures of mini-bursts is that one can probe much lower lens masses compared to what can be achieved via the lensing of the main burst.

We calculate how FRB lensing can probe compact objects if these objects have an extended mass distribution (Figs.\,\ref{fig:fDM_limit_extended_mass_distribution_MMin_1SolarMass} and \ref{fig:fDM_limit_extended_mass_distribution_MMax_1000SolarMass}).  We assume that the extended mass distribution follows a power law and assume various values of the power-law index and the maximum and minimum values of this extended mass distribution.  We find that in all cases, FRB lensing can probe constraints which are stronger than existing constraints.  In order to achieve these constraints, it is essential to detect a large number of FRBs with redshift information which near future radio telescopes can detect.

We also show that FRB lensing can detect exotic compact objects made up of beyond the Standard Model fermions or bosons.  In this case, the sizes of these compact objects are such that they can be lensed by FRB (Fig.\,\ref{fig:Upper limit on radius}).  It can be seen that FRB lensing is very efficient in constraining particle physics models in which the exotic particles can cluster.

The discovery of FRBs and gravitational wave are two major recent discoveries which can shed light on various major physics questions.  Lensing of these FRBs by compact objects can produce leading constraints over a wide range of lens mass.  This conclusion holds true for all the choices of the extended mass distributions of primordial black holes and the redshift distribution of FRBs that we studied.  Upcoming observations are expected to probe a large number of FRBs with redshift information, then FRB lensing will indeed prove to be one of the most powerful tools to constrain primordial black holes and other exotic compact objects.

%{\it Note added:}  After the publication of the first version of the paper, Refs.\,\cite{Choi:2019mva} and \cite{Sammons:2020kyk} studied probing bosons stars via FRBs and constraining compact dark matter using FRB microstructures respectively.  These two papers agree with our conclusions.  

%-----------------------------------------------------------------------------
\section*{Acknowledgments}
%-----------------------------------------------------------------------------

The author gratefully acknowledges helpful discussions with Julian B.\,Mu\~noz.  RL has received funding from
the DFG under Grants Nos.\ EXC-1098, FOR~2239,
GRK~1581, and from the European Research Council (ERC) under the European
Union's Horizon 2020 research and innovation programme (grant agreement No.\
637506, ``$\nu$Directions'') awarded to Joachim Kopp.

\bibliographystyle{kp}
\interlinepenalty=10000
\tolerance=100
\bibliography{refs}

%%%%%%%%%%%%%%%%%%%%%%%%%%%%%%%%%%%%%%%%%%%%%%%%%%%%%%%%
%%%%%%%%%%%%%%%%%%%%%%%%%%%%%%%%%%%%%%%%%%%%%%%%%%%%%%%%

\end{document}